\documentclass[longauth]{aa}  

\usepackage[switch]{lineno}
\usepackage{natbib}
\bibpunct{(}{)}{;}{a}{,}{,}
\usepackage{longtable}

\usepackage{graphicx,subcaption}
\usepackage{pifont,txfonts}

\usepackage{xspace}
\usepackage{pstricks}
\usepackage{color}
\graphicspath{{figures/}}

\usepackage{hyperref}
\hypersetup{
    unicode=true,           
    pdftoolbar=true,        
    pdfmenubar=true,        
    pdffitwindow=true,      
    pdftitle={Sylvia},
    pdfauthor={Carry et al.},
    pdfsubject={Planetary Science},
    pdfkeywords={},         
    pdfnewwindow=true,      
    colorlinks=true,        
    linkcolor=gray,         
    citecolor=blue,         
    filecolor=gray,         
    urlcolor=gray           
} 

\newcommand{\numb}[1]{\textcolor{orange}{#1}}
\renewcommand{\numb}[1]{#1}  

\newcommand{\adam}{\texttt{ADAM}\xspace}
\newcommand{\mpcd}{\texttt{MPCD}\xspace}
\newcommand{\genoid}{\texttt{Genoid}\xspace}
\newcommand{\mistral}{\texttt{Mistral}\xspace}

\newcommand{\rev}[1]{\textbf{#1}}
\renewcommand{\rev}[1]{#1}

\newcommand{\ObsOne}{\numb{143}}

\newcommand{\ObsTwo}{\numb{68}}

\newcommand{\rmsOrb}{\numb{9.3}}

\newcommand{\nbLC}{121}

\newcommand{\Diam}{\numb{271}}
\newcommand{\dDiam}{\numb{5}}

\newcommand{\Mass}{\numb{1.44}}
\newcommand{\dMass}{\numb{0.01}}
\newcommand{\eMass}{\numb{10$^{19}$}}

\newcommand{\Dens}{\numb{1378}}
\newcommand{\dDens}{\numb{45}}

\newcommand{\sid}{kg$\cdot$m$^{-3}$}
\newcommand{\jtwo}{J\ensuremath{_2}\xspace}
\newcommand{\jtwoval}{0.082}
\newcommand{\jtwounc}{0.005}

\usepackage{catoptions}
\makeatletter

\def\Autoref#1{%
  \begingroup
  \edef\reserved@a{\cpttrimspaces{#1}}%
  \ifcsndefTF{r@#1}{%
    \xaftercsname{\expandafter\testreftype\@fourthoffive}
      {r@\reserved@a}.\\{#1}%
  }{%
    \ref{#1}%
  }%
  \endgroup
}
\def\testreftype#1.#2\\#3{%
  \ifcsndefTF{#1autorefname}{%
    \def\reserved@a##1##2\@nil{%
      \uppercase{\def\ref@name{##1}}%
      \csn@edef{#1autorefname}{\ref@name##2}%
      \autoref{#3}%
    }%
    \reserved@a#1\@nil
  }{%
    \autoref{#3}%
  }%
}
\makeatother

\begin{document} 
    \title{Evidence for differentiation of the most primitive small bodies%
    \thanks{Based on observations made with 
      ESO Telescopes at the La Silla Paranal Observatory under program
      \href{http://archive.eso.org/wdb/wdb/eso/sched_rep_arc/query?progid=073.C-0851}{073.C-0851} (PI Merline),
      \href{http://archive.eso.org/wdb/wdb/eso/sched_rep_arc/query?progid=073.C-0062}{073.C-0062} (PI Marchis),
      \href{http://archive.eso.org/wdb/wdb/eso/sched_rep_arc/query?progid=085.C-0480}{085.C-0480} (PI Nitschelm),
      \href{http://archive.eso.org/wdb/wdb/eso/sched_rep_arc/query?progid=088.C-0528}{088.C-0528} (PI Rojo),
      \href{http://archive.eso.org/wdb/wdb/eso/sched_rep_arc/query?progid=199.C-0074}{199.C-0074} (PI Vernazza).},%
    \thanks{The reduced and deconvolved SPHERE images  are  available  at  the  CDS  via  anonymous   ftp  to \url{http://cdsarc.u-strasbg.fr/} or via \url{http://cdsarc.u-strasbg.fr/viz-bin/qcat?J/A+A/xxx/Axxx}}}
    
  \author{B.~Carry\inst{\ref{oca}}         \and 
    P.~Vernazza\inst{\ref{lam}}            \and 
    F.~Vachier\inst{\ref{imcce}}           \and 
    M.~Neveu\inst{\ref{maryland},\ref{goddard}}           \and 
    J.~Berthier\inst{\ref{imcce}}          \and 
    J.~Hanu{\v s}\inst{\ref{prague}}       \and 
    M.~Ferrais\inst{\ref{lam},\ref{liege}} \and 
    L.~Jorda\inst{\ref{lam}}               \and 
    M.~Marsset\inst{\ref{mit}}             \and 
    M.~Viikinkoski\inst{\ref{tampere}}     \and 
    P.~Bartczak\inst{\ref{poznan}}         \and 
    R.~Behrend\inst{\ref{geneva}}           \and 
    Z.~Benkhaldoun\inst{\ref{ouk}}         \and 
    M.~Birlan\inst{\ref{imcce},\ref{buca}}            \and 
    J.~Castillo-Rogez\inst{\ref{jpl}}      \and 
    F.~Cipriani\inst{\ref{estec}}          \and 
    F.~Colas\inst{\ref{imcce}}             \and 
    A.~Drouard\inst{\ref{lam}}             \and 
    G.~P.~Dudzi{\' n}ski\inst{\ref{poznan}}         \and 
    J.~Desmars\inst{\ref{imcce},\ref{ipsa}}      \and 
    C.~Dumas\inst{\ref{tmt}}               \and 
    J.~{\v D}urech\inst{\ref{prague}}      \and 
    R.~Fetick\inst{\ref{lam}}              \and 
    T.~Fusco\inst{\ref{lam}}               \and 
    J.~Grice\inst{\ref{oca},\ref{ou}}      \and 
    E.~Jehin\inst{\ref{liege}}             \and 
    M.~Kaasalainen\inst{\ref{tampere},\thanks{Deceased}}
        \and 
    A.~Kryszczynska\inst{\ref{poznan}}     \and 
    P.~Lamy\inst{\ref{lamy}}                \and 
    F.~Marchis\inst{\ref{seti}}            \and 
    A.~Marciniak\inst{\ref{poznan}}        \and 
    T.~Michalowski\inst{\ref{poznan}}      \and 
    P.~Michel\inst{\ref{oca}}              \and 
    M.~Pajuelo\inst{\ref{imcce},\ref{puc}} \and 
    E.~Podlewska-Gaca\inst{\ref{poznan},\ref{edyta}} \and 
    N.~Rambaux\inst{\ref{imcce}}           \and 
    T.~Santana-Ros\inst{\ref{tsr1}, \ref{tsr2}}      \and 
    A.~Storrs\inst{\ref{towson}}           \and 
    P.~Tanga\inst{\ref{oca}}               \and 
    A.~Vigan\inst{\ref{lam}}               \and 
    B.~Warner\inst{\ref{warner}}           \and 
    \rev{M.~Wieczorek}\inst{\ref{oca}}           \and 
    O.~Witasse\inst{\ref{estec}}           \and 
    B.~Yang\inst{\ref{eso}}                     
  }

   \institute{
     %
     Universit\'e C{\^o}te d'Azur, Observatoire de la
     C{\^o}te d'Azur, CNRS, Laboratoire Lagrange, France
     \email{benoit.carry@oca.eu}\label{oca}
     \and 
     Aix Marseille Univ, CNRS, LAM, Laboratoire d'Astrophysique de Marseille, Marseille, France
     \label{lam}
     \and     
     IMCCE, Observatoire de Paris, PSL Research University, CNRS, Sorbonne Universit{\'e}s, UPMC Univ Paris 06, Univ. Lille, France
     \label{imcce}
     \and 
     University of Maryland College Park, College Park, MD 20742, USA
     \label{maryland}
     \and
     NASA Goddard Space Flight Center, Greenbelt, MD 20771, USA
     \label{goddard}
     \and 
     Institute of Astronomy, Faculty of Mathematics and Physics, Charles University,
     V~Hole{\v s}ovi{\v c}k{\'a}ch 2, 18000 Prague, Czech Republic
     \label{prague}
     \and 
     Space sciences, Technologies and Astrophysics Research (STAR) Institute, Universit{\'e} de Li{\`e}ge, All{\'e}e du 6 Ao{\^u}t 17, 4000 Li{\`e}ge, Belgium
     \label{liege}
     %
     \and 
     Department of Earth, Atmospheric and Planetary Sciences, MIT, 77 Massachusetts Avenue, Cambridge, MA 02139, USA
     \label{mit}
     \and 
     Mathematics and Statistics, Tampere University, 33014 Tampere, Finland
     \label{tampere}
     \and 
     Astronomical Observatory Institute, Faculty of Physics, Adam Mickiewicz University, ul. S{\l}oneczna 36, 60-286 Pozna{\'n}, Poland
     \label{poznan}
     \and 
     Geneva Observatory, 1290 Sauverny, Switzerland
     \label{geneva}
     \and 
     Oukaimeden Observatory, High Energy Physics and Astrophysics Laboratory, Cadi Ayyad University, Marrakech, Morocco
     \label{ouk}
     \and 
     Astronomical Institute of the Romanian Academy, 5 Cutitul de Argint, 040557 Bucharest, Romania
     \label{buca}
     \and 
     Jet Propulsion Laboratory, California Institute of Technology, 4800 Oak Grove Drive, Pasadena, CA 91109, USA
     \label{jpl}
     \and 
     European Space Agency, ESTEC - Scientific Support Office, Keplerlaan 1, Noordwijk 2200 AG, The Netherlands
     \label{estec}
     \and 
     Institut Polytechnique des Sciences Avanc{\'e}es IPSA, 63 bis Boulevard de Brandebourg, F-94200 Ivry-sur-Seine, France\label{ipsa}
     \and 
     Thirty-Meter-Telescope, 100 West Walnut St, Suite 300, Pasadena, CA 91124, USA
     \label{tmt}
     \and 
     Open University, School of Physical Sciences, The Open University, MK7 6AA, UK
     \label{ou}
     \and 
     Laboratoire Atmosph\`eres, Milieux et Observations Spatiales, CNRS \& Universit\'e de Versailles Saint-Quentin-en-Yvelines, Guyancourt, France 
     \label{lamy}
     \and 
     SETI Institute, Carl Sagan Center, 189 Bernado Avenue, Mountain View CA 94043, USA 
     \label{seti}
     \and 
     Secci{\'o}n F{\'i}sica, Departamento de Ciencias, Pontificia Universidad Cat{\'o}lica del Per{\'u}, Apartado 1761, Lima, Per{\'u}
     \label{puc}
     \and 
     Institute of Physics, University of Szczecin, Wielkopolska 15, 70-453 Szczecin, Poland
     \label{edyta}
     \and 
     Departamento de F{\'i}sica, Ingenier{\'i}a de Sistemas y Teor{\'i}a de la Se{\~n}al, Universidad de Alicante, Alicante, Spain
     \label{tsr1}
     \and 
     Institut de Ci{\`e}ncies del Cosmos (ICCUB), Universitat de Barcelona (IEEC-UB), Mart{\'i} Franqu{\`e}s 1, E08028 Barcelona, Spain
    \label{tsr2}
     \and 
     Towson University, Towson, Maryland, USA
     \label{towson}
     \and 
     Center for Solar System Studies, 446 Sycamore Ave., Eaton, CO 80615, USA
     \label{warner}
     \and 
     European Southern Observatory (ESO), Alonso de Cordova 3107, 1900 Casilla Vitacura, Santiago, Chile
     \label{eso}
  }
  \date{Received September 15, 1996; accepted March 16, 1997}


  \abstract
   {Dynamical models of Solar System evolution have suggested that the so-called P- and D-type
   volatile-rich asteroids formed in the outer Solar System beyond Neptune's
   orbit and may be genetically related to the Jupiter Trojans, the comets and small Kuiper-belt objects (KBOs).
   \rev{Indeed}, the spectral properties of P/D-type asteroids 
   resemble that of anhydrous cometary dust.}
   {\rev{We aim at gaining insights into the above classes of bodies by characterizing} the internal structure of a large P/D-type asteroid.}
   {We report high-angular-resolution imaging observations of P-type asteroid (87) Sylvia with VLT/SPHERE.
   These images were used to reconstruct the 3D shape of Sylvia.
   Our images together with those obtained in the past with large ground-based telescopes were used
   to study the dynamics of its two satellites. We also model Sylvia's thermal evolution. 
   }
   {The shape of Sylvia appears flattened and elongated (a/b$\sim$1.45 ; a/c$\sim$1.84).
   We derive a volume-equivalent diameter of \Diam\,$\pm$\,\dDiam\,km, and a 
   low density of \Dens\,$\pm$\,\dDens\,\sid{}.
   The two satellites orbit Sylvia on circular, equatorial orbits.
   The oblateness of Sylvia should imply a detectable nodal precession which contrasts
   with the fully-Keplerian dynamics of its two satellites.
   This reveals an inhomogeneous internal structure,
   suggesting that Sylvia is differentiated.
   }
   {Sylvia's low density and differentiated interior can be explained by partial melting and 
   mass redistribution through water percolation.
   The outer shell would be composed of material similar to interplanetary dust particles (IDPs)
   and the core similar to aqueously altered IDPs or
   carbonaceous chondrite meteorites such as the Tagish Lake meteorite.
   Numerical simulations of the thermal evolution of Sylvia show that for a body of such size,
   partial melting was unavoidable due to the decay of long-lived radionuclides.
   In addition, we show that bodies as small as 130–150 km in diameter should have followed a similar thermal evolution, while smaller objects, such as comets and the KBO Arrokoth, must have remained pristine, in agreement with \textsl{in situ} observations of these bodies.
   NASA Lucy mission target (617) Patroclus (diameter $\approx$140\,km) may, however, be differentiated.}

  \keywords{Minor planets, asteroids: individual: Sylvia;  Kuiper belt: general}

\authorrunning{Carry et al.}
\titlerunning{Sylvia}
\maketitle


%
\section{Introduction}

  \indent The Cybele region, at the outer rim
  of the asteroid belt (3.27--3.7 au), is essentially populated by P- and D-type asteroids and
  to a lesser extent by C-type bodies
  \citep{2013Icar..226..723D,2014Natur.505..629D}.
  P- and D-type asteroids
  are thought to have formed in the outer Solar System (beyond 10\,au), among the progenitors
  of the current Kuiper Belt, and to have been implanted in the inner Solar System 
  (asteroid belt, Lagrangian Points of Jupiter) following
  giant planet migrations \citep[see][]{2009Natur.460..364L, 2014-Icarus-229-DeMeo, 2016-AJ-152-Vokrouhlicky}.
  This implies that the P/D-type main belt asteroids and the \rev{Jupiter Trojans} could be compositionally 
  related to outer small bodies such as Centaurs, short-period comets, 
  and small (D$\leq$300km) Kuiper Belt Objects (KBOs). 
  This dynamical scenario is currently supported by the similarity in size distributions
  between the Jupiter Trojans and small KBOs \citep{2014ApJ...782..100F} 
  as well as the similarity in terms of spectral properties between P/D-type main belt asteroids, 
  the Trojans of Jupiter and comets \citep{2015ApJ...806..204V, 2016arXiv161108731V}.
  
  \indent Overall, the outer Solar system is of tremendous interest, 
  as it is recognized as the least processed
  since the dawn of the Solar System and thus the closest to the \rev{primordial} materials
  from which the Solar System formed \citep[e.g.,][]{2020Sci...367.6620M}. 
  This is currently supported by the analysis of the spectral properties of P/D-type main belt asteroids, 
  \rev{Jupiter Trojans}, and comets that reveal a surface composition compatible with that of 
  anhydrous chondritic porous interplanetary dust particles
  \citep[CP IDP, see][]{2015ApJ...806..204V}.
  The CP IDPs are currently seen among the available extra-terrestrial materials as the
  closest to the starting ones \citep{1999ASIC..523..485B}. In particular, 
  \rev{based on its albedo and visible, near- and mid-infrared spectrum,}
  it is now well established
  \rev{\citep[see][]{2013Icar..225..517V}} 
  that the aqueously altered Tagish Lake meteorite cannot be representative
  of the surface composition of D-type asteroids nor that of most P-type asteroids
  as suggested earlier
  \citep{2001Sci...293.2234H}.
  \rev{As such, CP IDPs are currently the most likely analogs of the refractory materials present at the surface of P/D type asteroids.}

  \indent There are only four known large (diameter $\geq$100 km)  P/D-type asteroids with satellites:
  (87) Sylvia, (107) Camilla, (617) Patroclus, and (624) Hektor, with reported densities
  of
  1,400\,$\pm$\,200\,\sid{} \citep[e.g.,][]{2014Icar..239..118B},
  1,280\,$\pm$\,130\,\sid{} \citep{2018Icar..309..134P},
  800$^{+200}_{-100}$ to 1080\,$\pm$\,330\,\sid{}
  \citep{2006Natur.439..565M, 2010Icar..205..505M},
  and 
  1,000\,$\pm$\,300\,\sid{} \citep{2014ApJ...783L..37M}
  respectively.
  Patroclus is a double asteroid with nearly equally-sized
  150\,km-100\,km components
  \citep[e.g.,][]{2015AJ....149..113B, 2017AA...601A.114H, 2020Icar..35213990B}
  and is therefore atypical
  among asteroids but further strengthens the common origin of P/D asteroids and small KBOs
  \citep{2018NatAs...2..878N}, many of which likely formed as binaries
  \citep{2017NatAs...1E..88F, 2019NatAs...3..808N, 2020AA...643A..55R}. 
  Both Camilla and Sylvia are among the largest asteroids, with diameters
  above \numb{250} and \numb{280}\,km respectively \citep{2012PSS...73...98C}.
  Their low density implies a bulk composition that cannot consist only of refractory 
  materials but that must also comprise a large amount of volatiles \citep{2018Icar..309..134P}.

  \indent With an estimated diameter of nearly \numb{280}\,km \citep[][and reference therein]{2012PSS...73...98C},
  (87) Sylvia is the largest body in the Cybele region and more generally the largest P/D-type asteroid in the inner ($\leq$5.5 au) Solar System. Its surface composition is fully consistent with that of anhydrous chondritic porous IDPs
  \citep{2015ApJ...806..204V, 2019PASJ...71....1U}. 
  Furthermore, it is the first asteroid around which two satellites were discovered \citep{2005Natur.436..822M}.
  Because of its rather large angular size at opposition ($\approx$0.14\arcsec) and its two moons, Sylvia is an ideal target for high angular-resolution adaptive-optics observations as these allow an accurate characterization of its 3D shape and of the mass of the system, hence of its bulk density. Furthermore, the two moons allow probing the internal structure and in particular the harmonics of the gravity field (at least the lowest-order gravitational moment, the quadrupole \jtwo). This information, in turn, probes the distribution of material inside Sylvia, i.e., whether its interior is homogeneous or differentiated.

  \indent Several authors have studied Sylvia's dynamical system
  \citep{2005Natur.436..822M, 2009MNRAS.395..218W, 2012-Icarus-220-Frouard,
  2012AJ....144...70F, 2014Icar..241...13B, 2014Icar..239..118B, 2016Icar..276..107D}.
  While these studies agreed on the mass of Sylvia and its low density (around 1,300\,\sid{}),
  there is no consensus regarding its gravitational potential
  (of which only the quadrupole \jtwo has been studied, \Autoref{tab:sum87}).
  The latter is, however, a direct consequence of the internal structure of Sylvia.
  
  To constrain the bulk density of Sylvia and its internal structure,
  we observed it as part of our imaging survey of D$\geq$100\,km main-belt
  asteroids 
  \citep[ID \href{http://archive.eso.org/wdb/wdb/eso/abstract/query?ID=9900740&progid=199.C-0074(A)}{199.C-0074}, PI P. Vernazza, see][]{%
  2018AA...618A.154V, 
  2020NatAs...4..136V, 
  2018AA...619L...3V,  
  2019AA...623A...6F,  
  2019AA...623A.132C, 
  2019AA...624A.121H, 
  2020AA...633A..65H, 
  2020NatAs...4..569M, 
  2020AA...641A..80Y, 
  2020AA...638L..15F}. 
  We imaged Sylvia over two apparitions,
  separated by a year, throughout its rotation at high
  angular resolution with the 
  SPHERE extreme adaptive-optics (AO) 
  IRDIS and ZIMPOL cameras
  \citep{2008SPIE.7014E..3FT, 2008SPIE.7014E..3LD, 2019AA...631A.155B}
  mounted on the European Southern Observatory (ESO)
  Very Large Telescope (VLT).
  Beyond contributing to our understanding of the formation and evolution of the Solar System's
  most primitive bodies, the present study provides the context for the future \textsl{in-situ}
  exploration of primitive P/D-type bodies by NASA's \textsl{Lucy} mission to the Jupiter Trojans.

  \begin{table*}[ht]
    \begin{center}
      \begin{tabular}{lc l@{\,$\pm$\,}l l@{\,$\pm$\,}l l@{\,$\pm$\,}l l}
        \hline\hline
        Satellite & N\textsubscript{obs} &
        \multicolumn{2}{c}{Mass} & 
        \multicolumn{2}{c}{Density} &
        \multicolumn{2}{c}{\jtwo} & Reference\\
        && \multicolumn{2}{c}{($\times 10^{19}$\,kg)} & 
        \multicolumn{2}{c}{(\sid)} &
        \multicolumn{2}{c}{} & \\
        \hline
        Romulus & 24    & 1.48  & 0.01  & 1200 & 100 & 0.17       & 0.05       &\citet{2005Natur.436..822M}\\
        Remus   & 12    & 1.47  & 0.01  & 1200 & 100 & 0.18       & 0.01       &\citet{2005Natur.436..822M}\\
        Both    & 45+20 & 1.484 & \rev{0.015} & 1290 & 390 & 0.099\,59  & 0.000\,84  &\citet{2012AJ....144...70F}\\
        Both    & 51+17 & 1.476 & 0.006 & 1200 & 100 & 0.000\,002 & 0.000\,300 &\citet{2014Icar..241...13B}\\
        Romulus & 65    & 1.476 & 0.166 & 1380 & 150 & 0          & 0.01       &\citet{2014Icar..239..118B}\\
        Remus   & 25    & 1.380 & 0.223 & 1290 & 200 & 0          & 0.024      &\citet{2014Icar..239..118B}\\
        Romulus & 76    & 1.470 & 0.008 & 1350 &  40 & \multicolumn{2}{c}{--}  &\citet{2016Icar..276..107D}\\
        Both    & \ObsOne+\ObsTwo & \Mass & \dMass & \Dens & \dDens & 0 & 0.01 & Present study\\
        \hline
      \end{tabular}
      \caption{Mass, density, and quadrupole \jtwo{} of Sylvia, derived using either Romulus, Remus, or both satellite,
      from the literature compared with the present study.}
      \label{tab:sum87}
    \end{center}
  \end{table*}

\section{Observations and data reduction\label{sec:obs}}

  \subsection{High-angular-resolution imaging\label{sec:obs:ao}}

    \indent The data used in the present study to extract the position of the moons
    comprises all the high angular resolution images of Sylvia 
    taken with the Hubble Space Telescope (HST) and 
    ground-based telescopes equipped with adaptive-optics (AO) cameras:
    Gemini North, ESO VLT, W. M. Keck, and
    SOR \citep{2016Icar..276..107D}.
    The data span \numb{51} different epochs, with
    multiple images each, over \numb{17} years from February 2001 to
    November 2018. For the reconstruction of Sylvia's 3D shape, however,
    only the images with the highest resolution were used, i.e.,
    those acquired with VLT/SPHERE/ZIMPOL (\Autoref{sec:shape}).
    
    \indent The images from the HST were obtained with
    the second Wide Field and Planetary Camera 
    \citep[WFPC2,][]{1995PASP..107..156H}.
    The images from the VLT were acquired with
    both the first generation instrument NACO
    \citep[NAOS-CONICA,][]{2003-SPIE-4841-Lenzen,2003-SPIE-4839-Rousset}
    and 
    SPHERE
    \citep[Spectro-Polarimetric High-contrast Exoplanet REsearch,][]{2006-OExpr-14-Fusco,2019AA...631A.155B}, the
    second generation extreme-AO instrument designed for 
    exoplanet detection and characterization.
    The images acquired with SPHERE were taken with both the IRDIS 
    \citep[InfraRed Dual-band Imager and Spectrograph,][]{2008SPIE.7014E..3LD}
    and ZIMPOL 
    \citep[Zurich Imaging Polarimeter,][]{2008SPIE.7014E..3FT}
    sub-systems.
    Images taken at the Gemini North used the NIRI camera
    \citep[Near InfraRed Imager,][]{2003-PASP-115-Hodapp}, fed by the ALTAIR AO system \citep{2000-AOST-Herriot}.
    Finally, observations at Keck were
    acquired with the guiding camera of NIRSPEC in 2001 \citep{1998SPIE.3354..566M}
    and NIRC2
    \citep[Near-InfraRed Camera 2,][]{2004-AppOpt-43-vanDam,2000-SPIE-4007-Wizinowich} later on.

    \indent To reduce the AO-imaging data, a standard data processing protocol (sky subtraction, bad-pixel
    identification and correction, and flat-field 
    correction) was followed using
    in-house routines developed in Interactive Data Language (IDL) 
    \citep[see][]{2008AA...478..235C}.
    The images were then deconvolved with the \mistral~algorithm 
    \citep{2000-PhD-Fusco, 2004-JOSAA-21-Mugnier}
    to restore their optimal angular resolution
    \citep[see][for details]{2019AA...623A...6F}. 
    Separately, the reduced images were processed to subtract the bright halo surrounding Sylvia
    to enhance the detectability of the satellites
    \citep[see][for details]{2018Icar..309..134P, 2019AA...623A.132C}.

  \subsection{Optical lightcurves\label{sec:obs:lc}}

    \indent We used the \numb{40} lightcurves from 
    \citet{2002Icar..159..369K} to create a convex 3-D shape model of
    Sylvia\footnote{Available on DAMIT \citep{2010-AA-513-Durech}:\\
      \href{https://astro.troja.mff.cuni.cz/projects/damit/}{https://astro.troja.mff.cuni.cz/projects/damit/}},
    compiled from the Uppsala Asteroid Photometric
    Catalog\footnote{\href{http://asteroid.astro.helsinki.fi/apc/asteroids/}{http://asteroid.astro.helsinki.fi/apc/asteroids/}}
    \citep{PDS-SAPC-2011}.
    We also compiled \numb{11} lightcurves acquired by amateur astronomers within the Courbes de rotation d'ast{\' e}ro{\" i}des et de com{\` e}tes database (CdR\footnote{\href{http://obswww.unige.ch/~behrend/page_cou.html}{http://obswww.unige.ch/\textasciitilde behrend/page\_cou.html}}).

    \indent In addition to these data, we acquired \numb{12}
    lightcurves using the  
    60\,cm \textsl{Andr{\'e} Peyrot} telescope mounted at Les Makes
    observatory on La R{\'e}union Island, operated as a 
    partnership among Les Makes Observatory and the IMCCE, Paris
    Observatory, and
    \numb{7} lightcurves with the 60\,cm TRAPPIST telescopes
    located at La Silla Observatory in Chile and the Ouka{\"i}meden observatory in 
    Morocco \citep{2011Msngr.145....2J}.
    Finally, we extracted \numb{51} lightcurves from the data archive of
    the SuperWASP survey \citep{2006-PASP-118-Pollacco} for the
    period 2006-2009 \citep{2005-EMP-97-Parley, 2017-ACM-Grice}.
    In summary, a total of \numb{\nbLC} lightcurves observed between
    \numb{1978} and \numb{2017}
    (\Autoref{tab:lc}) were used in this work and are presented in 
    \Autoref{fig:app:lc}.

  \subsection{Stellar occultations\label{sec:obs:occ}}

    \indent \numb{Nine} stellar occultations by Sylvia have been recorded
    since 1984, mostly by amateur astronomers during the last decade
    \citep[see][]{2014-ExA-38-Mousis, 2016-IAU-Dunham, 2020MNRAS.499.4570H}.
    We converted the timings of disappearance and reappearance of the
    occulted stars\footnote{compiled by \citet{PDSSBN-OCC} and publicly available on the Planetary Data
    System (PDS): \href{http://sbn.psi.edu/pds/resource/occ.html}{http://sbn.psi.edu/pds/resource/occ.html}}
    into segments (called chords) on the plane
    of the sky, using the location of the observers on Earth and the
    apparent motion of Sylvia following the recipes listed in \citet{1999-IMCCE-Berthier}.
    Only five stellar occultations had multiple chords that could be used to constrain
    the size and apparent shape of Sylvia (\Autoref{fig:app:occ}).
    For the January 2013 and October 2019 occultations, several observers
    reported secondary events due to
    the occultation of the stars by either Romulus or Remus
    \citep{2014Icar..239..118B, 2019-CBET-Vachier}.
    We thus used the relative positions between
    Sylvia and its satellites at the time of the occultations
    to constrain their mutual orbits.
    We list the observers of the occultations in \Autoref{tab:occ}.

\section{Sylvia's 3D shape\label{sec:shape}}

  \indent We fed the All-Data Asteroid Modeling (\adam{})
  algorithm with all SPHERE/ZIMPOL images, lightcurves,
  and stellar occultations to determine the spin and 3D shape of Sylvia
  \citep{2015-AA-576-Viikinkoski}.
  Optical lightcurves are often required for \adam{} to 
  constrain the regions not imaged and to 
  stabilize the shape optimization.
  The procedure is similar to that published in our previous studies with SPHERE,
  and we refer to these for more details
  \citep[e.g.,][]{2015-AA-581-Viikinkoski, 2017AA...604A..64M, 2018AA...618A.154V}.
  We used the sidereal rotation period and the
  spin-axis coordinates of Sylvia from the literature
  \citep{2002Icar..159..369K,2013Icar..226.1045H,2017AA...601A.114H} as
  input values to \adam{}.
  
  \indent We further model the shape by using the Multi-resolution PhotoClinometry by Deformation
  (\mpcd{}) method \citep{2016Icar..277..257J}, following the procedure of our previous works
  \citep[e.g.,][]{2020AA...638L..15F}.
  \mpcd gradually deforms the vertices of a previous mesh (here \adam{} model)
  to minimize the difference between
  the observed images and  synthetic images of the model
  \citep{2010SPIE.7533E..11J}.
  Both models only present marginal differences (\Autoref{tab:shape}),
  and in what follows, we report on the \mpcd model. 
  
  \indent The derived shape is in essence similar to that based
  on lower angular resolution images
  \citep{2014Icar..239..118B, 2017AA...601A.114H}.
  A striking feature of Sylvia is its remarkable elongated shape (\Autoref{tab:shape},
  \Autoref{fig:elong}). To put its peculiar shape into context,
  we measured the tri-axial diameters of \numb{103} asteroids larger than 100\,km
  from their shape models\footnote{Retrieved from DAMIT \citep{2010-AA-513-Durech}:\\
      \href{https://astro.troja.mff.cuni.cz/projects/damit/}%
      {https://astro.troja.mff.cuni.cz/projects/damit/}}
  and compiled their rotation periods from the Planetary Data System \citep{2017-PDS-Harris}. Sylvia appears to be more elongated and to spin faster than most asteroids larger than
  100\,km (\Autoref{fig:elong}).
  In particular, the population of asteroids with satellites stands out from the population
  of singletons, 
  with a median ratio of equatorial diameters $a/b$ of \numb{1.37 (\textsl{vs} 1.16} for the background)
  and a median
  rotation period of \numb{5.5\,h (\textsl{vs} 7.9\,h} for the background).
  We compared the $a/b$ ratio and rotation period distributions of 
  asteroids with and without satellites using the Kolmogorov-Smirnov test.
  The p-values for both are \numb{$5 \cdot 10^{-4}$} and 
  \numb{$7 \cdot 10^{-5}$} respectively.
  We thus conclude that the distributions are different above the 99.5\% confidence level. 
  \rev{While asteroids with a diameter smaller than about 15\,km are subject to YORP
  spin-up and surface re-arrangment
  \citep{2008Natur.454..188W, 2015-AsteroidsIV-Vokrouhlicky},
  Sylvia is too large to have been affected.
  The origin of this difference is thus unclear.}
  It may result from the impact at the origin of the satellite formation
  \citep{2015-AsteroidsIV-Margot}, or alternatively,
  satellites may be more stable around elongated bodies \citep[see][]{2009MNRAS.395..218W}.

\begin{figure*}[ht]
    \centering
    \includegraphics[width=.99\hsize]{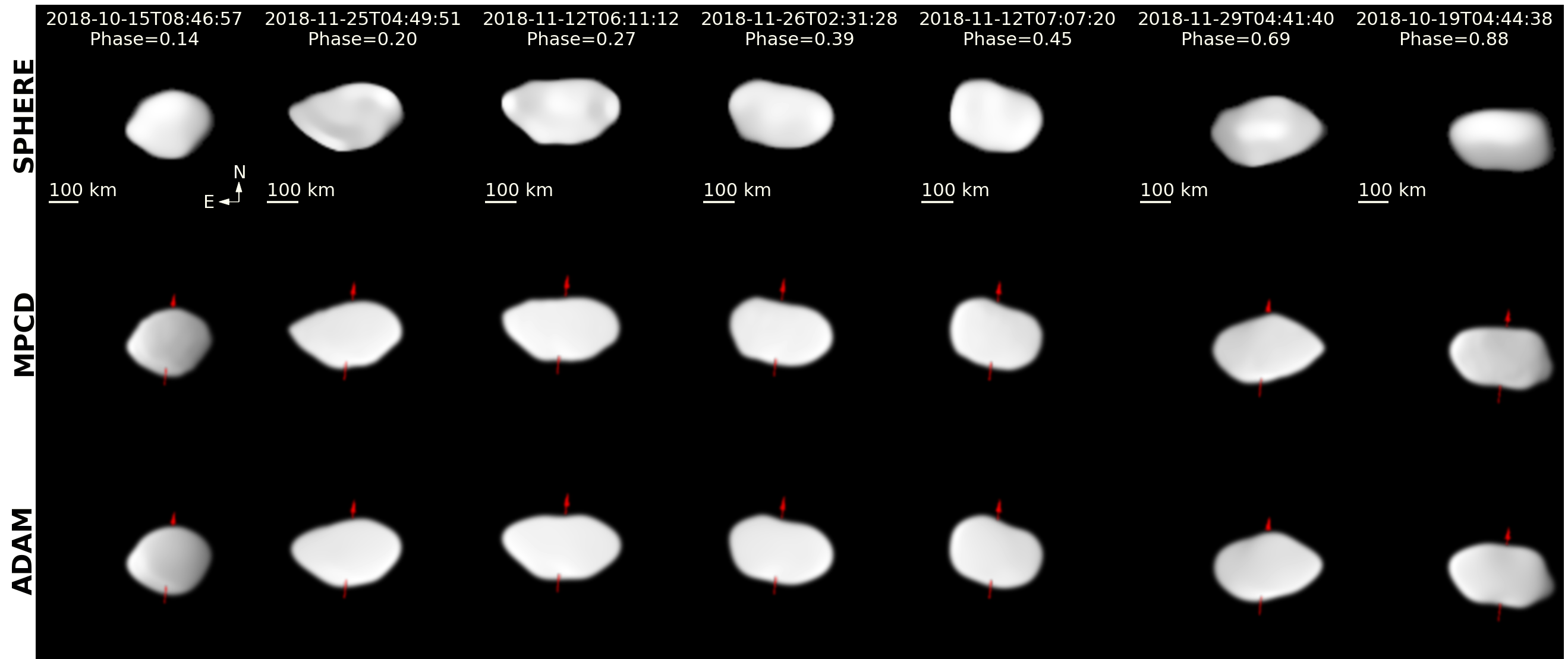}
    \caption{Comparison of the SPHERE image (top row) with the
    MPCD (middle) and ADAM (bottom) models.}
    \label{fig:shape_ao}
\end{figure*}

\begin{figure}[ht]
    \centering
    \includegraphics[width=0.99\hsize]{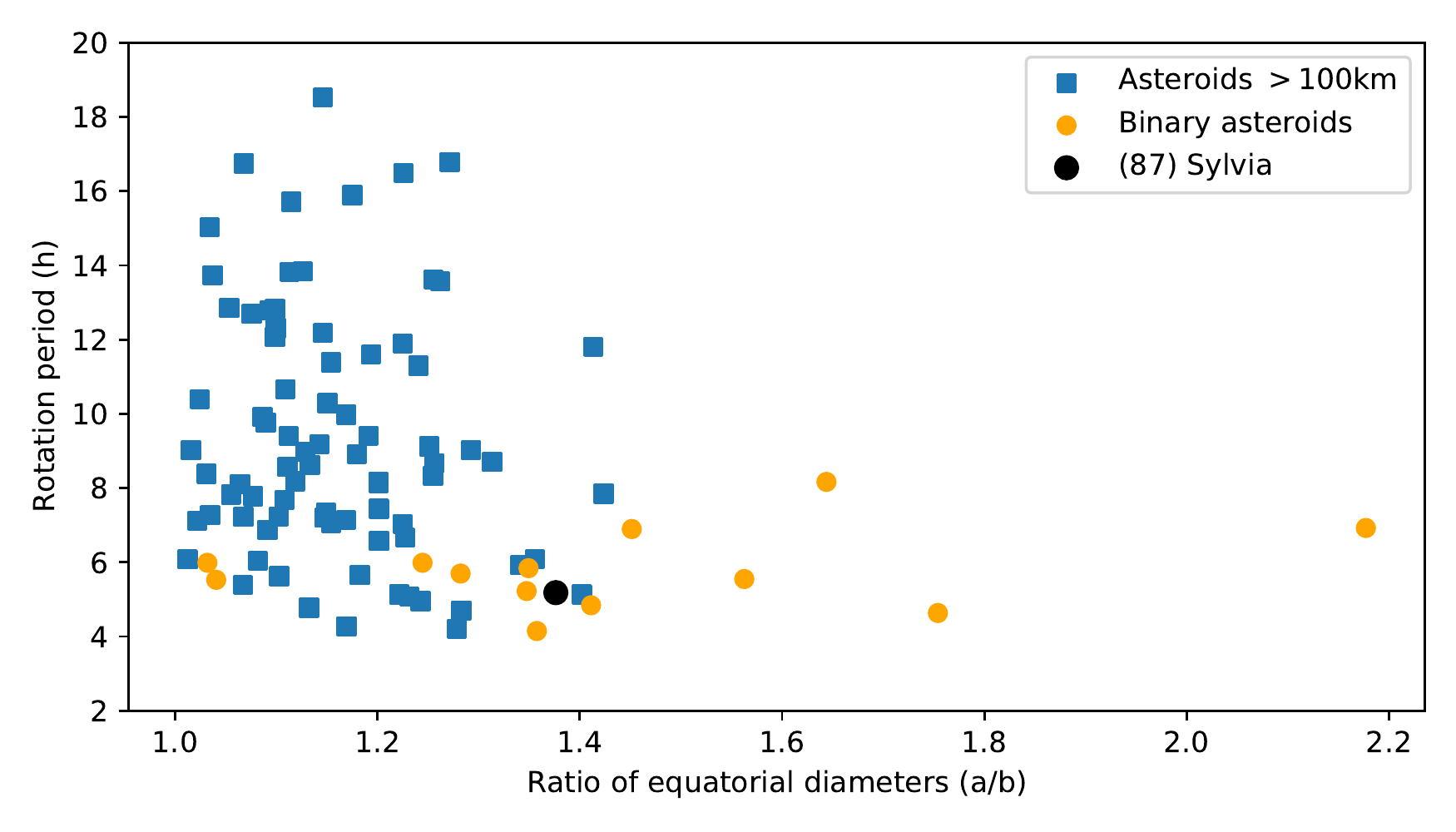}
    \caption{Distribution of the ratio of equatorial diameters ($a/b$)
    and rotation period of \numb{103}
    asteroids larger than 100\,km in diameter.
    The difference between asteroids with and without satellites is striking.}
    \label{fig:elong}
\end{figure}

\begin{table*}[t]
\begin{center}
  \caption{Spin solution (coordinates in ecliptic and
      equatorial J2000 reference frames) and shape model parameters
      (the overall shape is reported as the $a>b>c$ diameters of a
      triaxial ellipsoid fit to the shape model). All uncertainties
    are reported at 1\,$\sigma$.
    \label{tab:shape}
  }
  \begin{tabular}{llllll}
    \hline\hline
    \multicolumn{2}{l}{Parameter} & \mpcd{} & \adam{} & Unc. & Unit \\
    \hline 
    Sidereal period & $P_s$     & \multicolumn{2}{c}{5.183\,641} & 3.9 $\cdot 10^{-5}$ & hour \\
    Longitude       &$\lambda$  & \multicolumn{2}{c}{ 75.3} & 5 & deg. \\
    Latitude        & $\beta$   & \multicolumn{2}{c}{+64.2} & 5 & deg. \\
    Right Ascension & $\alpha$  & \multicolumn{2}{c}{ 14.3} & 5 & deg. \\
    Declination     & $\delta$  &\multicolumn{2}{c}{ +83.5} & 5 & deg. \\
    Ref. epoch      & T$_0$     & \multicolumn{2}{c}{2443750.000} & \\
    \hline
    Diameter   & $\mathcal{D}$   & \Diam & 274 & \dDiam & km \\
    Volume     & V   &  1.05 $\cdot 10^{7}$ & 1.08 $\cdot 10^{7}$ & 2 $\cdot 10^{5}$ & km$^3$ \\
    Diameter a & a   & 363 & 374 & 5 & km\\
    Diameter b & b   & 249 & 248 & 5 & km\\
    Diameter c & c   & 191 & 194 & 5 & km\\
    Axes ratio & a/b & 1.46 & 1.51 & 0.03 & \\
    Axes ratio & b/c & 1.30 & 1.28 & 0.04 & \\
    Axes ratio & a/c & 1.90 & 1.93 & 0.05 & \\
    \hline
  \end{tabular}
\end{center}
\end{table*}



\section{Dynamics of the system\label{sec:dyn}}

  \indent The two satellites orbit Sylvia on equatorial, circular,
  and prograde orbits (\Autoref{tab:dyn}).
  The root mean square (RMS) residual between the observations
  and the computed positions is only \numb{\rmsOrb}\,mas,
  that is within the pixel size of most observations.
  The positive occultations by both satellites
  in October 2019 provide a practical estimate
  of the reliability of the orbital solution, as the two satellites were detected
  at only \numb{5\,mas} from the positions we predicted
  \citep{2019-CBET-Vachier}.
  It also highlights the importance of accurate ephemerides to prepare the 
  observation of the occultation by placing observers on the path of the satellites.
  
\begin{figure}[ht]
    \centering
    \includegraphics[width=\hsize]{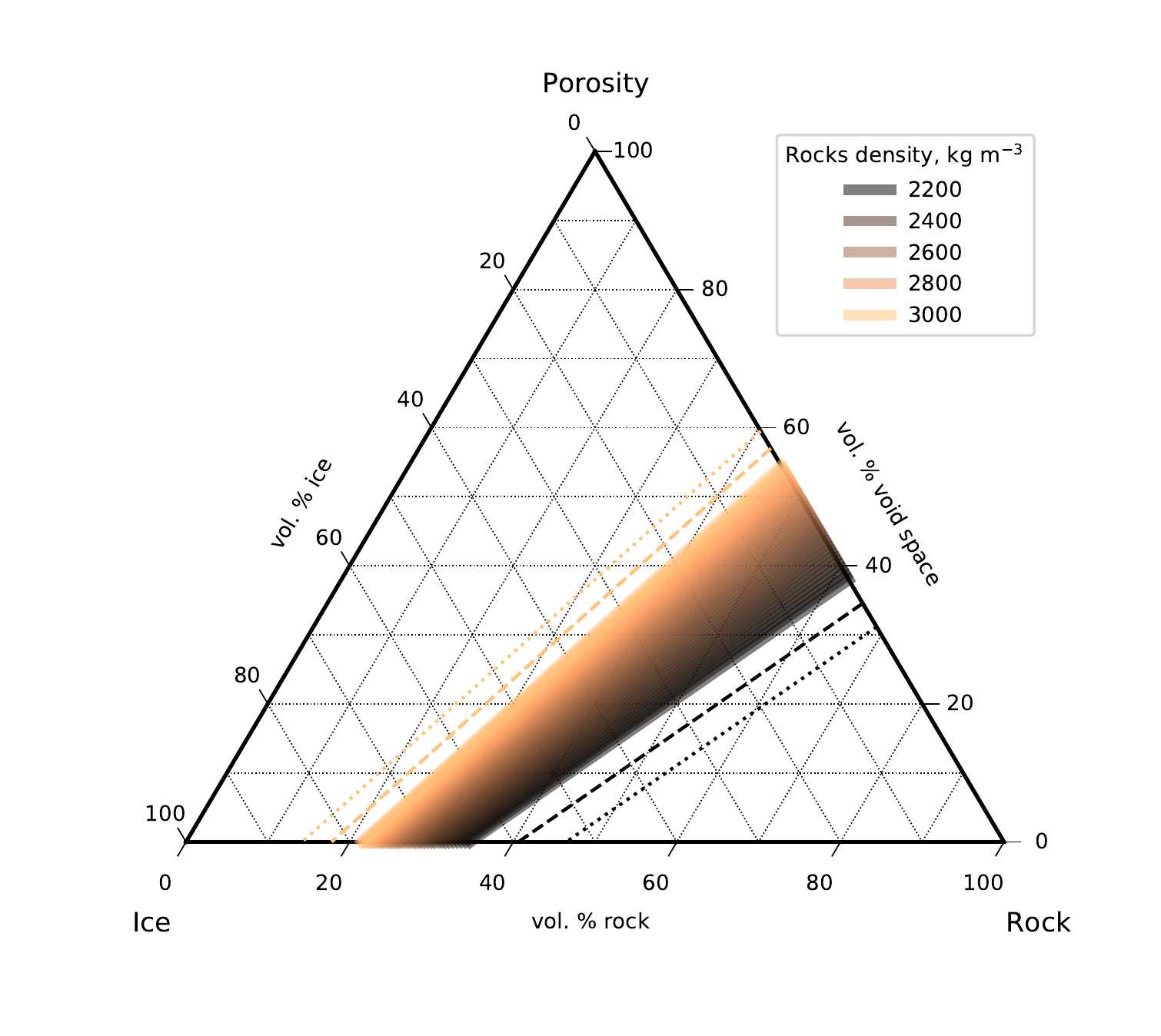}
    \caption{Bulk composition of Sylvia, assuming three end members: rocks, ices (density of 920\,\sid{}),
    and voids. The dashed and dotted lines represent the 1 and 2\,$\sigma$
    boundaries.}
    \label{fig:ternary}
\end{figure}

  \indent The mass of Sylvia is constrained with an uncertainty of less than 1\% : 
  (\Mass\,$\pm$\,\dMass) $\times$ \eMass\,kg, thanks to the long baseline of observations.
  Combined with our volume-equivalent diameter estimate (\Diam\,$\pm$\,\dDiam\,km, see above),
  the density of Sylvia is found to be
  \Dens\,$\pm$\,\dDens\,\sid{}, reminiscent of that of other
  large asteroids with a surface composition consistent with that of IDPs
  \citep[C/P/D types, see][]{2012PSS...73...98C, 2015ApJ...806..204V}.
  We present in \Autoref{fig:ternary} all the possible bulk compositions
  of Sylvia, considering a mixture of rocks and ices, with voids.
  We use a range of density from 2200 to 3000\,\sid{}
  for the rock phase \citep[from rocks with organics to the 
  density of the silicate phase reported by the \textit{Stardust} mission,][]{2006Sci...314.1711B}.
  The density of Sylvia implies the presence of both ices and macroporosity
  in its interior.

  \indent We estimate the masses of Romulus and Remus at
  \numb{(1.4\,$\pm$\,1.2)\,$\times$\,10$^{15}$}\,kg and
  \numb{(7.8\,$\pm$\,7.3)\,$\times$\,10$^{14}$}\,kg
  (i.e., effectively upper limits), 
  very close to the masses reported by \citet{2012AJ....144...70F}.
  Assuming a similar albedo for Sylvia and its two moons, 
  their magnitude differences with Sylvia (\Autoref{tab:dyn}) imply
  diameters of \numb{15$_{-6}^{+10}$\,km and 10$_{-6}^{+17}$\,km} for Romulus and Remus
  \citep[consistent with occultation chords,][]{2014Icar..239..118B}, hence densities of 
  \numb{790\,$\pm$\,680}
  and
  \numb{1480\,$\pm$\,1400}\,\sid{}.
  The density of both satellites is loosely constrained and similar to that of Sylvia.
  
\begin{figure}[ht]
    \centering
    \includegraphics[width=.99\hsize]{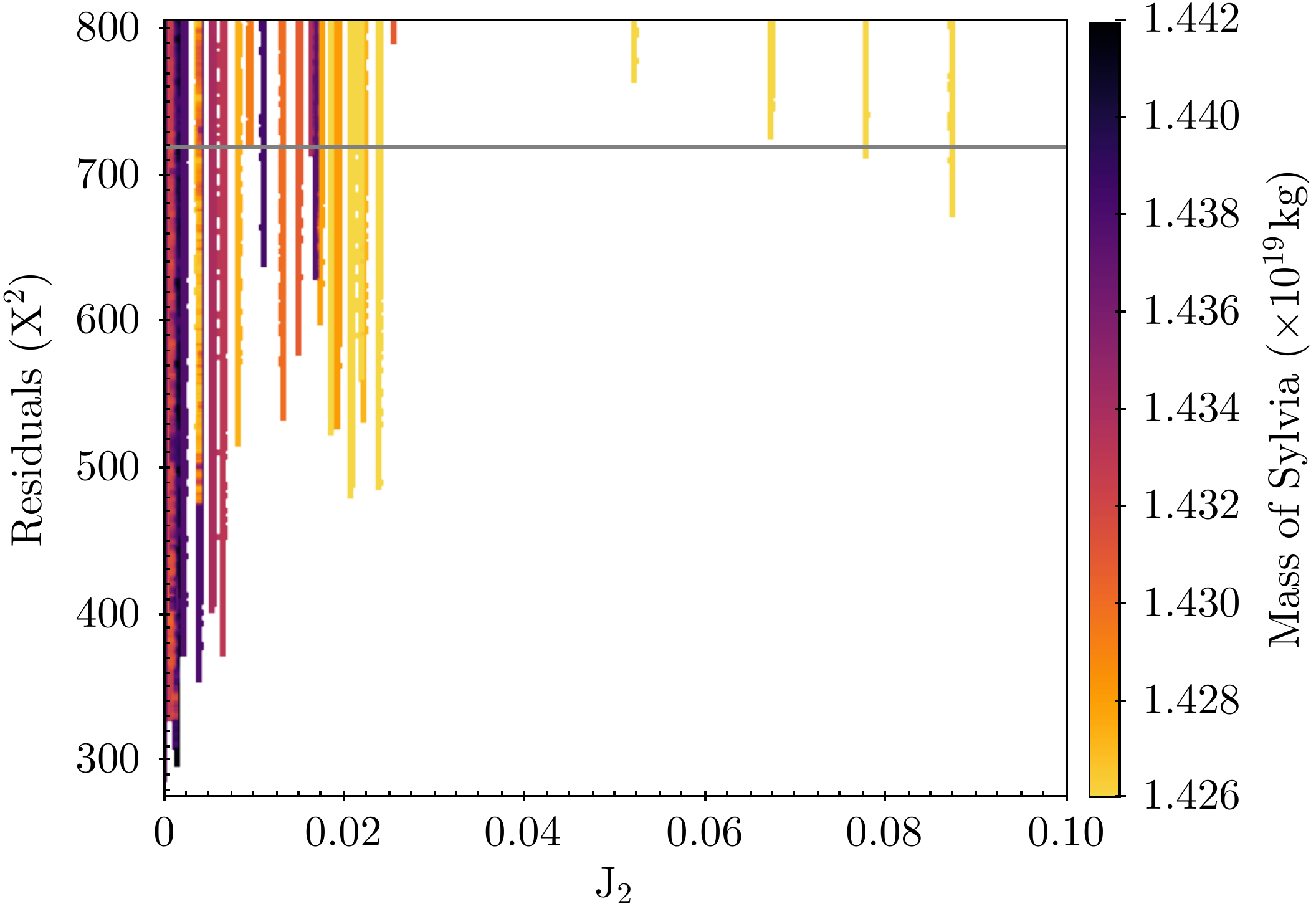}
    \caption{Orbital residuals ($\chi^2$) as function
    of the dynamical quadrupole \jtwo. The horizontal grey line corresponds
    to the $\chi^2$ providing a fit at 1\,$\sigma$ of the observations.}
    \label{fig:j2}
\end{figure}

  \indent Finally, we note that the best orbital solutions
  are obtained for the smallest
  quadrupole \jtwo (\Autoref{fig:j2}), tending toward \jtwo=0 
  (i.e., fully Keplerian orbits over the 19 years baseline).
  Although there are orbits fitting the data within 1\,$\sigma$ of the
  observations, their residuals are systematically larger.

\begin{table*}
\begin{center}
  \caption[Orbital elements of the satellites of Sylvia]{%
    Orbital elements of the satellites of Sylvia,
    expressed in EQJ2000, obtained with \genoid:
    orbital period $P$, semi-major axis $a$,
    eccentricity $e$, inclination $i$,
    longitude of the ascending node $\Omega$,
    argument of pericenter $\omega$, time of pericenter $t_p$.
    The number of observations and RMS between predicted and
    observed positions are also provided.
    Finally, we report the mass of Sylvia $M_{\textrm{Sylvia}}$,
    the mass of Romulus $M_{\textrm{Romulus}}$,
    the mass of Remus $M_{\textrm{Remus}}$,
    their apparent magnitude difference $\Delta m$ with Sylvia,
    the ecliptic J2000 coordinates of the orbital pole
    ($\lambda_p,\,\beta_p$), 
    the equatorial J2000 coordinates of the orbital pole
    ($\alpha_p,\,\delta_p$), and the
    orbital inclination ($\Lambda$) with respect to the equator of
    Sylvia. Uncertainties are given at 3-$\sigma$.}
  \label{tab:dyn} 
   \begin{tabular}{l ll ll}
    \hline\hline
    & \multicolumn{2}{c}{Romulus} & \multicolumn{2}{c}{Remus}\\ 
    \hline
  \noalign{\smallskip}
  \multicolumn{2}{c}{Observing data set} \\
  \noalign{\smallskip}
    Number of observations  & \multicolumn{2}{c}{130} & \multicolumn{2}{c}{66} \\ 
    Time span (days)        & \multicolumn{2}{c}{6050} & \multicolumn{2}{c}{5656} \\ 
    RMS (mas)               & \multicolumn{2}{c}{9.85} & \multicolumn{2}{c}{8.24} \\ 
    \hline
  \noalign{\smallskip}
  \multicolumn{2}{c}{Orbital elements EQJ2000} \\
  \noalign{\smallskip}
    $P$ (day)         & 3.64126 & $\pm$ 0.00005 & 1.35699 & $\pm$ 0.00075 \\ 
    $a$ (km)          & 1340.6 & $\pm$ 1.2 & 694.2 & $\pm$ 0.4 \\ 
    $e$               & 0.000 & $_{-0.000}^{+0.009}$ & 0.005 & $_{-0.005}^{+0.031}$ \\ 
    $i$ (\degr)       & 7.4 & $\pm$ 1.6 & 8.7 & $\pm$ 5.4 \\ 
    $\Omega$ (\degr)  & 97.1 & $\pm$ 5.8 & 100.8 & $\pm$ 20.6 \\ 
    $\omega$ (\degr)  & 171.0 & $\pm$ 10.5 & 262.2 & $\pm$ 25.9 \\ 
    $t_{p}$ (JD)      & 2455597.08689 & $\pm$ 0.10085 & 2455594.89253 & $\pm$ 0.10444 \\ 
    \hline
  \noalign{\smallskip}
  \multicolumn{2}{c}{Physical parameters} \\
  \noalign{\smallskip}
    $M_{\textrm{Sylvia}}$ ($\times 10^{19}$ kg)      & 1.440 & $\pm$ 0.004 \\ 
    $M_{\textrm{Romulus}}$ ($\times 10^{15}$ kg)     & 1.4 & $\pm$ 1.2 \\ 
    $M_{\textrm{Remus}}$ ($\times 10^{14}$ kg)      & 7.8 & $\pm$ 7.3 \\ 
    $\Delta m_{\textrm{Romulus}}$       & 6.2 & $\pm$ 1.1 \\ 
    $\Delta m_{\textrm{Remus}}$         & 7.1 & $\pm$ 2.1 \\ 
    \hline
  \noalign{\smallskip}
  \multicolumn{2}{c}{Derived parameters} \\
  \noalign{\smallskip}
    $\lambda_p,\,\beta_p$ (\degr)      & 73, +65 & $\pm$ 4, 1 & 70, +64 & $\pm$ 11, 3 \\ 
    $\alpha_p,\,\delta_p$ (\degr)      & 7, +83 & $\pm$ 6, 2 & 11, +81 & $\pm$ 21, 5 \\ 
    $\Lambda$ (\degr)                  & 5 & $\pm$ 1 & 6 & $\pm$ 4 \\ 
    $\mathcal{D}_{\textrm{Romulus}}$ (km) & 15.1 & $\pm$ 1.1 \\ 
    $\mathcal{D}_{\textrm{Remus}}$   (km) & 10.3 & $\pm$ 2.1 \\ 
    \hline
  \end{tabular}
\end{center}
\end{table*}

\section{Implication for the internal structure\label{sec:disc}}

  \indent Under the assumption of a homogeneous density in the interior,
  the shape of Sylvia implies a \jtwo of \numb{\jtwoval\,$\pm$\,\jtwounc}
  \citep[computed with
  \texttt{SHTOOLS}\footnote{\href{https://shtools.oca.eu}{https://shtools.oca.eu}},
  see][]{2018GGG....19.2574W}.
  This values contrasts with the null \jtwo determined dynamically
  (\Autoref{sec:dyn}).
  This discrepancy reveals an inhomogeneous density distribution inside Sylvia
  and
  hints at a more spherical mass concentration than suggested by
  Sylvia's oblate and elongated 3D shape.
  This implies a denser, more spherical \textsl{core},
  surrounded by a less dense \textsl{envelope}.
  Based on similar considerations, similar internal structures have
  recently been proposed
  for other large P-types such as the Cybele (107) Camilla \citep{2018Icar..309..134P}
  and the Jupiter Trojan (624) Hektor \citep{2014ApJ...783L..37M}.
 
  
\begin{figure*}[ht]
    \centering
    \includegraphics[width=.99\hsize]{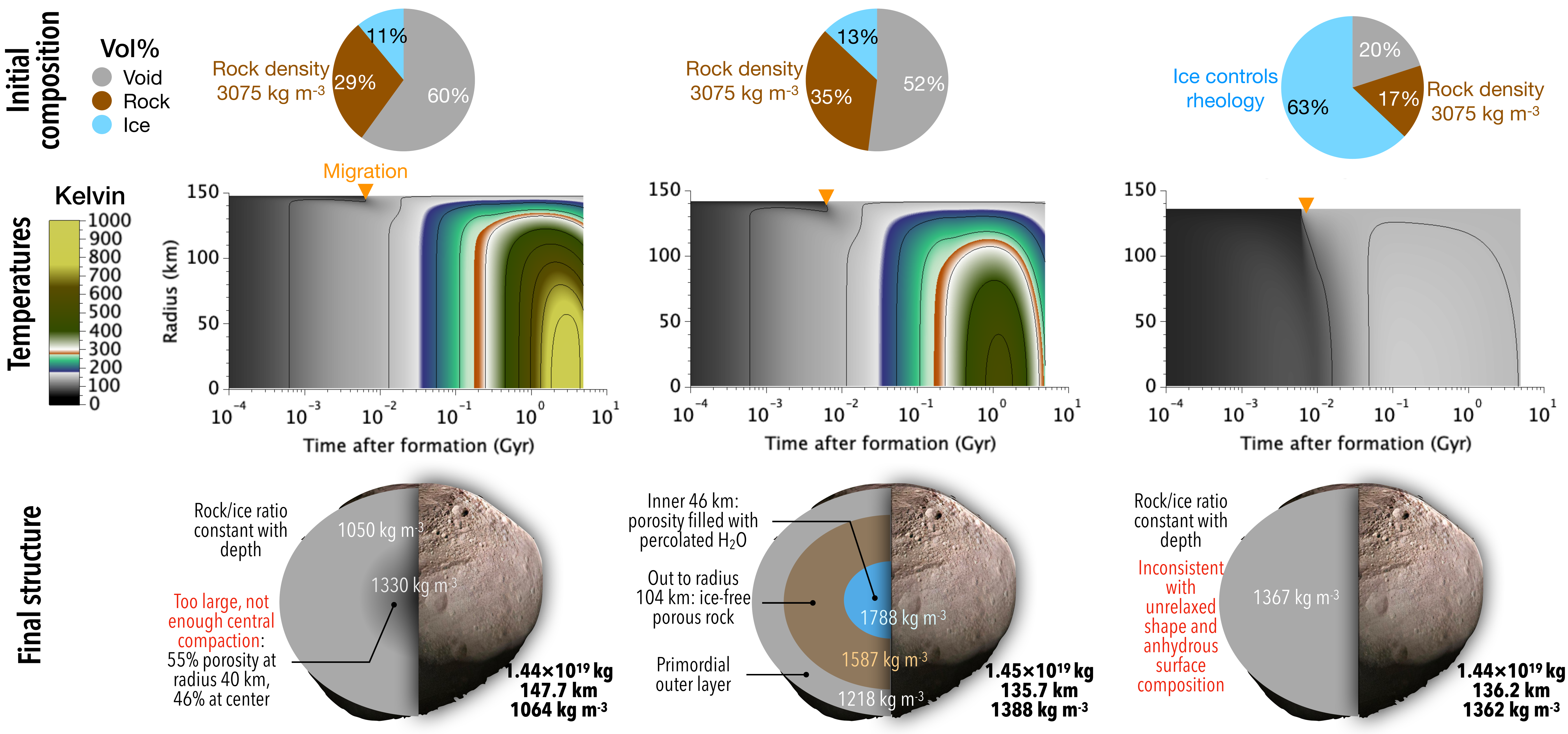}
    \caption{%
    Long-term evolution of the internal structure of Sylvia.
    The baseline scenario is presented in the central column, while the
    left and right columns present extreme cases in which the structure is dictated by rock-compaction and water ice rheology, respectively.
    }
    \label{fig:evol}
\end{figure*}
  
  This differentiated structure is at odds with the IDP-like spectral properties,
  evidence for an absence of both thermal metamorphism and aqueous alteration.
  This suggests that partial differentiation occurred, limited by the insufficient
  amount of heat generated by radionuclides which did not propagate to the 
  surface.
  Such structures have indeed been suggested for the parent bodies of 
  CV carbonaceous chondrites \citep{2011EPSL.305....1E},
  ordinary chondrites \citep{2019JGRE..124.1880B},
  and mid-sized KBOs \citep{desch2009thermal}.
  
  \indent Building upon the work of \citet{2019ApJ...875...30N},
  we model the thermal and internal structure history of Sylvia.
  The evolution of internal temperatures and structure is computed numerically using a one-dimensional code \citep{desch2009thermal}.
  Sylvia is assumed to be made of rock (idealizing a mixture of refractory materials such as silicates, metals, and organic material),
  water ice, and voids (the macroporosity).
  \rev{The mass is distributed assuming spherical symmetry 
  over 200 grid zones initially evenly spaced in radius.
  The internal energy in each grid zone is computed from the initial 
  temperature using equations of state for rock and ice.
  Material is never hot enough in our simulations for rock-metal differentiation,
  which is neglected. Initial radionuclide abundances are provided in 
  Table 1 of \citet{2019ApJ...875...30N}. 
  Simulations start once Sylvia is fully formed, neglecting the progressive accretion of material over time.
  Because of this and the near-absence of short-lived radionuclide heating 
  given the assumed formation time, Sylvia's simulated early evolution is cold. 
  The implementation of instantaneous differentiation in the central regions 
  that warm above 273\,K rests on the assumption that sufficiently large 
  rock grains settle via Stokes flow on timescales smaller than one time step. }
  
  \indent \rev{Sylvia} is assumed to accrete homogeneously at 60 K, 6 million years (My) after the formation of Ca-Al-rich inclusions
  \citep[consistent with a surface without aqueous alteration; ][]{2019ApJ...875...30N}.
  It is the equilibrium temperature for an albedo 0.05 at a distance of 17-18 au \citep[i.e.,
  the postulated accretion distance of KBOs,][]{morbidelli2020kuiper} from the Sun with
  70\% of the present-day luminosity.
  The surface temperature is instantaneously raised to 148\,K (the present-day equilibrium temperature)
  at the time of heliocentric migration. We test different timings,
  from a late planet migration (hundreds of My) such as the one described
  by the Nice model
  \citep{2005Natur.435..459T, 2005Natur.435..462M,
         2005Natur.435..466G, 2009Natur.460..364L}
  to an early dynamical instability occurring a few My after the dissipation of
  the gas disk \citep{2018NatAs...2..878N, 2019AJ....157...38C},
  and find that the timing of implantation into the asteroid belt seldom affects its
  structure or peak temperature.
  In the results below, the time of migration is set to 6 My after formation
  (i.e., 12 My after Ca-Al-rich inclusions).
  
\begin{figure}[ht]
    \centering
    \includegraphics[width=\hsize]{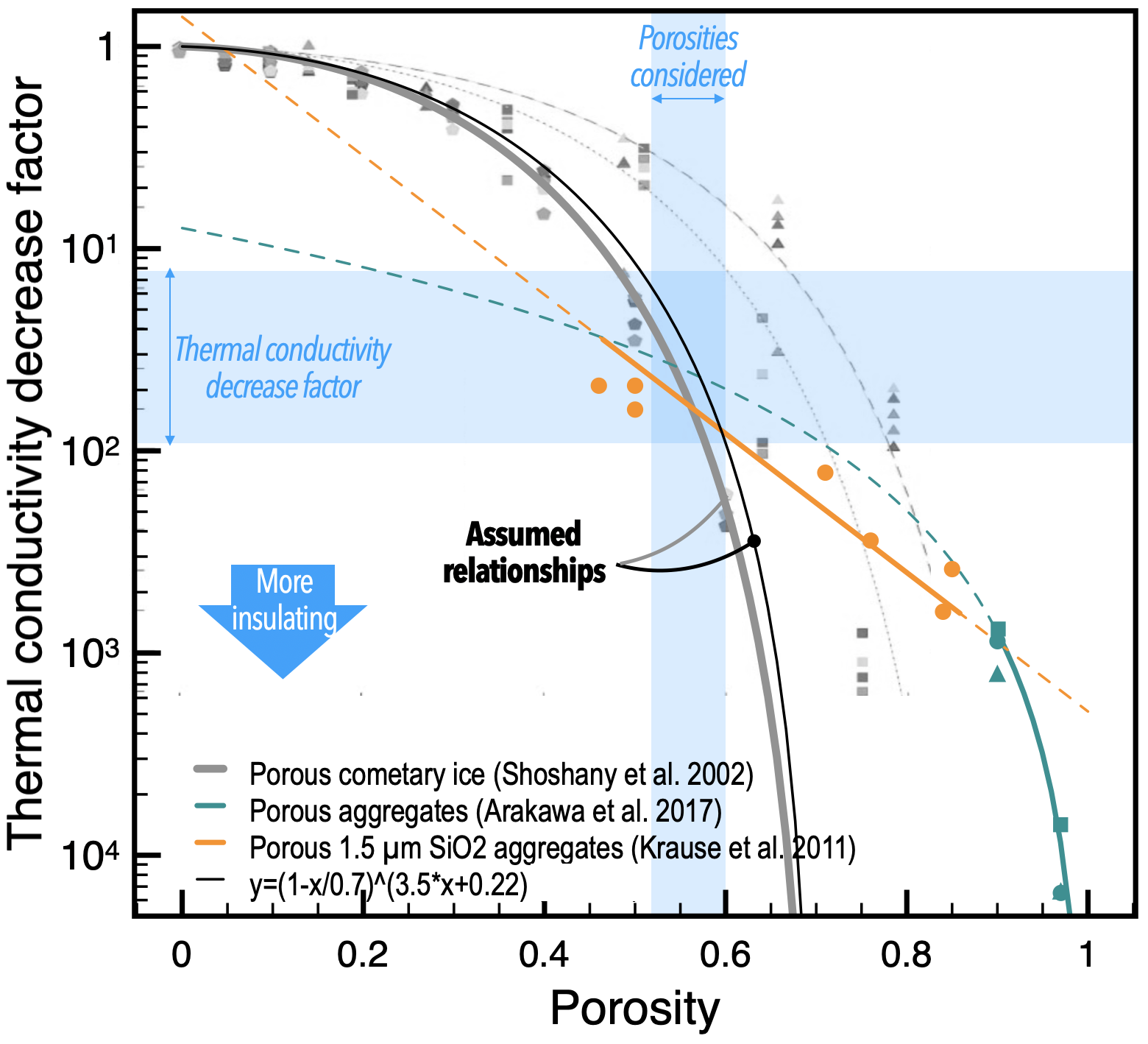}
    \caption{Effect of porosity $\Phi$ on the thermal conductivity $k$. The orange and teal curves show relationships that were only validated in the regimes where the curves are solid. The orange curve is a thermal conductivity in W m$^{-1}$ K$^{-1}$ (rather than a decrease factor). Data points of a given color are from the same source as the fit curve of that color.}
    \label{fig:app:kporo}
  \end{figure}

  \indent The thermal structure is determined by balancing conductive heat transfer with primarily
  radiogenic heating by $^{26}$Al, $^{40}$K, $^{232}$Th, $^{235}$U, and $^{238}$U,
  using a finite-difference method and a 50-yr time step, for 5 billion years (Gy).
  Thermal conductivities depend mainly on porosities (\Autoref{fig:app:kporo}),
  but also on composition and temperature \citep{desch2009thermal}.
  Porosity is allowed to compact at rates determined from material viscosities as described in
  \citet{neveu2017origin}.
  Sylvia's bulk density constrains the void porosity and rock volume fractions to about 40--55\% and 20--35\%,
  respectively, mainly depending on the rock density (\Autoref{fig:ternary}). 
   
  \rev{Convection is generally neglected as it is assumed that the postulated porous, 
  rock-rich internal structure for Sylvia is not prone to fluid or ice advection. 
  In the ice-rich case (\Autoref{fig:evol}, right panel), 
  solid-state convection is allowed to occur but does not, 
  because the critical Rayleigh number is never exceeded. 
  Although this was not simulated, in the percolation case below we expect convection 
  to be possible in the central region rich in liquid water, 
  until this region refreezes. In all simulations, 
  volume changes due to water melting or freezing are neglected.}
   
  \rev{The viscosity of ice-rock mixtures, 
  used to compute the Rayleigh number and pore compaction, 
  is calculated following \citet{roberts2015fluffy}. 
  Above 30 vol.\% ice, it is equated to the viscosity of ice. 
  Below 30\% ice volume fraction, as a first approximation, 
  it is set to the geometric mean of the rock and ice viscosities at given grain size, 
  stress (equated with hydrostatic pressure), and temperature. 
  \citet{roberts2015fluffy} noted that this approximation tends 
  to underestimate mixture viscosities relative 
  to extrapolations of laboratory measurements. 
  The ice and rock flow laws adopted in the model are the composite rheology of 
  \citet{goldsby2001superplastic} and the dry diffusion creep flow law 
  for olivine of \citet{korenaga2008new}, respectively.}
  
  \rev{The key factor governing thermal evolution in these simulations is porosity $\Phi$, 
  which decreases thermal conductivities $k$ of rock-ice mixtures, 
  of order 1 W m$^{-1}$ K$^{-1}$ \citep[and references therein]{desch2009thermal}, 
  by up to two orders of magnitude. 
  The adopted thermal conductivity-porosity relationship \citep{shoshany2002monte} is derived from 
  Monte-Carlo modeling of porous cometary ice: 
  $k$ is decreased with increasing $\Phi$ via multiplication by a factor 
  (1 - $\Phi$ /0.7)$^{n\Phi+0.22}$. 
  This relationship is shown in \Autoref{fig:app:kporo}. 
  We adopt $n$=4.1 (grey curve) for the canonical simulation with 52\% porosity, 
  following the discussion in \citet{shoshany2002monte}. 
  We set $n$=3.5 (black curve) for the simulation with 60\% porosity; 
  choosing $n$=4.1 would result in more heating and compaction than shown in \Autoref{fig:evol}. 
  There is a wide spread in the Monte Carlo results of \citet{shoshany2002monte} in how 
  $\Phi$ affects $k$, with some of their models suggesting a lesser effect of porosity. 
  Conversely, independent measurements of porous silica aggregates \citep{krause2011thermal} 
  and extrapolated results from models of highly porous aggregates \citep{arakawa2017thermal} 
  both suggest similar or slightly higher decrease factors due to porosity 
  (\Autoref{fig:app:kporo}) than the relationships we have assumed. 
  Thus, the assumed relationships seem to adequately represent the state 
  of understanding of how porosities of 50 to 60 vol.\% decrease thermal conductivities. }

  Crucially, to be compatible with Sylvia's observed anhydrous surface and low $J_2$ despite an oblate shape,
  time-evolution simulations with this model constrain the volume fraction of water to be low, less than 40 vol\% relative to rock
  or 15 vol\% overall.
  For higher volume fractions, ice grains tend to become adjoined and control the mechanical properties of the interior.
  In that case, the material viscosity is assumed equal to that of water ice \citep{goldsby2001superplastic},
  and any void porosity rapidly decreases as the interior warms due to radiogenic heating and porosity insulation.
  Instantaneous ice-rock differentiation happens first once the central (warmest) regions warm above 273\,K.
  It then proceeds outward if the interior keeps warming.
  This yields a gravitationally unstable structure: the topmost undifferentiated layers are denser than underlying layers,
  which are poorer in rock. Once Sylvia is differentiated out to more than half its radius,
  differentiation is assumed to proceed by gravitational (Rayleigh-Taylor) instabilities:
  layers overturn if their viscosity is below a threshold that corresponds to T$\approx$\rev{150}\,K \citep{rubin2014effect}.
  Since Sylvia's post-migration surface temperature is warmer, 148\,K,
  differentiation out to the surface is \rev{essentially} inevitable.
  This ought to result in evidence of surface water, either as ice or as mineral hydration,
  as observed on asteroids linked to carbonaceous chondrites \citep{2010-Nature-464-Rivkin, 2010-Nature-464-Campins}.
  This is inconsistent with Sylvia's anhydrous, IDP-like surface composition.
  Ice-rock differentiation can be prevented if ice dominates the volume fraction (\Autoref{fig:evol}, right column),
  since in that case the combination of low rock (i.e., radionuclide) content and low insulating void porosity results
  in a cold interior in which ice never melts.
  However, in such a homogeneous interior, the mass distribution should
  result in
  a higher $J_2$ than observed given Sylvia's oblate shape.
  
  It follows that to retain a pristine anhydrous external envelope, Sylvia's water volume fraction must be
  low \citep[consistently with observations of the comet 67P nucleus by the \textit{Rosetta} spacecraft; ][]{patzold2019nucleus,choukroun2020dust}
  so that interior solids are less prone to deformation, inhibiting both porosity compaction and instability-driven differentiation.
  As a canonical case, we assume an interior comprised of 52 vol.\% void porosity, 35\% rock of density 3075 kg m$^{-3}$, and 13\% water
  (\Autoref{fig:evol}, central column).
  The void porosity decreases the interior's thermal conductivity by a factor of $\approx$15 relative to
  a compact rock-ice mixture. This favors accumulation of radiogenic heat, melting ice in the central regions after
  $\approx$0.15-0.2~Gy. At such low water volume fraction and high void porosity,
  it is sensible for liquid water to percolate downward in pores without significantly disturbing the remaining
  rock-void porosity structure, since rock grains already tend to be adjoined
  \citep[see Fig. 3a,b of][for a pictorial description]{neumann2020ceres}.
  This would result in a three-layer structure (\Autoref{fig:evol}, bottom central panel):
  a central region where the porosity has been filled with percolated water, surrounded by a porous layer 
  free of water, and a primordial outer layer that remains too cold for ice to melt. 
  Our thermal evolution simulations do not explicitly track
  percolation. The example interior structure of \Autoref{fig:evol},
  bottom central panel is obtained by manually moving the mass of water that is liquid at 0.2\,Gy 
  to fill porosity at the center, and assuming that in the middle layer, the empty volume left behind 
  by the displaced water is compacted (so that the void porosity remains 52\% in this layer).
  This compaction leads to a volume-averaged diameter decrease from 283.4\,km to 271.4\,km, the observed value.
  The spherical central mass distribution in this three-layer model implies a \jtwo{} of \numb{$5 \cdot 10^{-5}$}
  only, consistent with the observed dynamics of
  both satellites.

  Although this is not a unique solution, assuming lower rock densities (\Autoref{fig:ternary}) requires
  increasing the rock volume fraction at the expense of ice or void porosity so as to keep matching Sylvia's
  bulk density. Neither the bulk ice volume fractions nor the bulk void porosity are likely to be much lower
  than assumed given the need to invoke the migration of melted ice, enabled by the insulating effect of porosity,
  to explain a more spherical mass distribution (low $J_2$) than suggested by Sylvia's oblate shape. 
  
  Another, less likely explanation for a spherical mass distribution inside Sylvia is central compaction of the rock.
  Although we assume a rather low viscosity for rock-ice mixtures,
  Sylvia's relatively low gravity (lithostatic pressures) precludes compaction below 900 K.
  The required thermal insulation could be achieved with a bulk void porosity as low as 60\% (\Autoref{fig:evol}; left column).
  However, such a hot evolution would result in advection and, likely, outward outgassing of
  water \citep{prialnik1999changes, young2003conditions}, 
  which are not captured in these simulations and would cool the interior.
  
  We thus deem percolation of water in the deep interior as being the likelier explanation for Sylvia's
  low $J_2$ despite its oblate shape. This implies that, unlike the pristine outer layers comparable 
  to the CP IDPs, the innermost region may be instead analogous to hydrated material exemplified by
  chondritic smooth IDPs or perhaps the Tagish Lake meteorite \citep{2019NatAs...3..910F}. 
  The minimum body diameter for such percolation to take place (holding all other quantities constant)
  is between 130 and 150\,km, implying that even objects as small as 
  Patroclus \citep[diameter $\approx$140 km\rev{, see}][]{2017AA...601A.114H}, target of NASA's \textit{Lucy} mission,
  may have experienced a low degree of central liquid water percolation.

\section{Conclusions}

  We used newly acquired high-angular resolution imaging observations of (87) Sylvia with
  the SPHERE instrument on the ESO VLT, along with archival images, lightcurves,
  and stellar occultations to reconstruct its 3D shape and to constrain
  the orbital properties of its two moons.
  
  We find that Sylvia possesses a low density of \Dens\,$\pm$\,\dDens\,\sid{}, similar to that of other
  large C/P/D asteroids whose surface composition is mostly consistent with that
  of anhydrous interplanetary dust particles.
  Sylvia spins fast and is oblate and elongated, a property shared by most 100+\,km
  multiple asteroids, contrasting with the physical properties of large asteroids without satellites.
  
  The orbits of the two satellites is in apparent 
  contradiction with the oblate shape of Sylvia: 
  the two orbits do not show the nodal precession expected from the shape. 
  We interpret it as an evidence for a central spherical mass concentration, due to water percolation
  over millions of years triggered by long-lived radionuclides.
  This long lasting heating episode allowed for partial \textsl{differentiation},
  the outer shell of Sylvia remaining pristine.
  It follows that even the most primitive small bodies with diameters larger than 150 km
  did not avoid thermal processing leaving only their outermost layers intact.

\begin{acknowledgements}

  Some of the work presented here is
  based on observations collected at the European Organisation
  for Astronomical Research in the Southern Hemisphere under ESO
  programs
  \href{http://archive.eso.org/wdb/wdb/eso/sched_rep_arc/query?progid=073.C-0851}{073.C-0851}
  (PI Merline),
  \href{http://archive.eso.org/wdb/wdb/eso/sched_rep_arc/query?progid=073.C-0062}{073.C-0062}
  (PI Marchis),
  \href{http://archive.eso.org/wdb/wdb/eso/sched_rep_arc/query?progid=085.C-0480}{085.C-0480}
  (PI Nitschelm),
  \href{http://archive.eso.org/wdb/wdb/eso/sched_rep_arc/query?progid=088.C-0528}{088.C-0528}
  (PI Rojo),
  \href{http://archive.eso.org/wdb/wdb/eso/sched_rep_arc/query?progid=199.C-0074}{199.C-0074}
  (PI Vernazza).\\

  \indent Some of the data presented herein were obtained at the W.M.
  Keck Observatory, which is operated as a scientific partnership
  among the California Institute of Technology, the University of
  California and the National Aeronautics and Space
  Administration. The 
  Observatory was made possible by the generous financial support
  of the W.M. Keck Foundation.\\
  
  \indent  This research has made use of the Keck Observatory Archive
  (KOA), which is operated by the W. M. Keck Observatory and the
  NASA Exoplanet Science Institute (NExScI), under contract with the
  National Aeronautics and Space Administration.\\

  \indent The authors wish to recognize and acknowledge the very significant cultural role and reverence that the summit of Mauna Kea
  has always had within the indigenous Hawaiian community. We
  are most fortunate to have the opportunity to conduct observations
  from this mountain.\\
  
  \indent We thank the AGORA association which administrates the
  60 cm telescope at Les Makes observatory, La Reunion island, under a financial
  agreement with Paris Observatory. Thanks to A. Peyrot, J.-P. Teng
  for local support, and A. Klotz for helping with the robotizing.\\

  \indent B. Carry, P. Vernazza, A. Drouard, and J. Grice were supported by CNRS/INSU/PNP.
  This work has been supported by the Czech Science Foundation
  through grants 20-08218S  (J. Hanu{\v s}, J. {\v D}urech) and
  by the Charles University Research program No. UNCE/SCI/023. 
  The work of TSR was carried out through grant APOSTD/2019/046 by Generalitat Valenciana (Spain).
  This work was supported by the MINECO (Spanish Ministry of Economy) through grant RTI2018-095076-B-C21 (MINECO/FEDER, UE). \\

  \indent \rev{Our colleague and co-author M. Kaasalainen passed away while
    this work was carried out. Mikko's influence in asteroid 3D shape modeling
    has been enormous. This study is dedicated to his memory.}\\

  \indent This paper makes use of data from the DR1 of the WASP data
  \citep{2010-AA-520-Butters} as provided by the WASP consortium, 
  and the computing and storage facilities at the CERIT Scientific
  Cloud, reg. no. CZ.1.05/3.2.00/08.0144 
  which is operated by Masaryk University, Czech Republic. 
  TRAPPIST-South is funded by the Belgian Fund for Scientific Research
  (Fond National de la Recherche Scientifique, FNRS) under the grant 
  \rev{PDR T.0120.21}, with the participation of the Swiss FNS. TRAPPIST-North is a
  project funded by the University of Liège, and performed in collaboration with
  Cadi Ayyad University of Marrakesh.
  E. Jehin is a Belgian FNRS Senior Research Associate.\\

  \indent Thanks to all the amateurs worldwide who regularly observe
  asteroid lightcurves and stellar occultations. The great majority
  of observers have made these observations at their own expense,
  including occasions when they have travelled significant distances.
  Most of those observers are affiliated with one or more of
  (i) European Asteroidal Occultation Network (EAON),
  (ii) International Occultation Timing Association (IOTA),
  (iii) International Occultation Timing Association – European Section (IOTA–ES),
  (iv) Japanese Occultation Information Network (JOIN),
  (v) Trans Tasman Occultation Alliance (TTOA). \\

  \indent The authors acknowledge the use of the Virtual Observatory
  tools
  Miriade\,\footnote{Miriade: \href{http://vo.imcce.fr/webservices/miriade/}{http://vo.imcce.fr/webservices/miriade/}}
  \citep{2008-ACM-Berthier}, 
  TOPCAT\,\footnote{TOPCAT:
    \href{http://www.star.bris.ac.uk/~mbt/topcat/}{http://www.star.bris.ac.uk/~mbt/topcat/}}, and
  STILTS\,\footnote{STILTS: \href{http://www.star.bris.ac.uk/~mbt/stilts/}{http://www.star.bris.ac.uk/~mbt/stilts/}}
  \citep{2005ASPC..347...29T}. This research used the
  SSOIS\,\footnote{SSOIS:
    \href{http://www.cadc-ccda.hia-iha.nrc-cnrc.gc.ca/en/ssois}{http://www.cadc-ccda.hia-iha.nrc-cnrc.gc.ca/en/ssois}}
  facility of the Canadian Astronomy Data Centre operated by the
  National Research Council of Canada with the support of the Canadian
  Space Agency \citep{2012-PASP-124-Gwyn}.
  Thanks to the developers and maintainers.

\end{acknowledgements}

\bibliographystyle{aa} 
\bibliography{current} 

\begin{appendix}


\section{Lightcurve observations\label{sec:app:lc}}

\begin{figure*}[h]
  \centering
  \begin{subfigure}{\textwidth}
    \includegraphics[width=.9\linewidth]{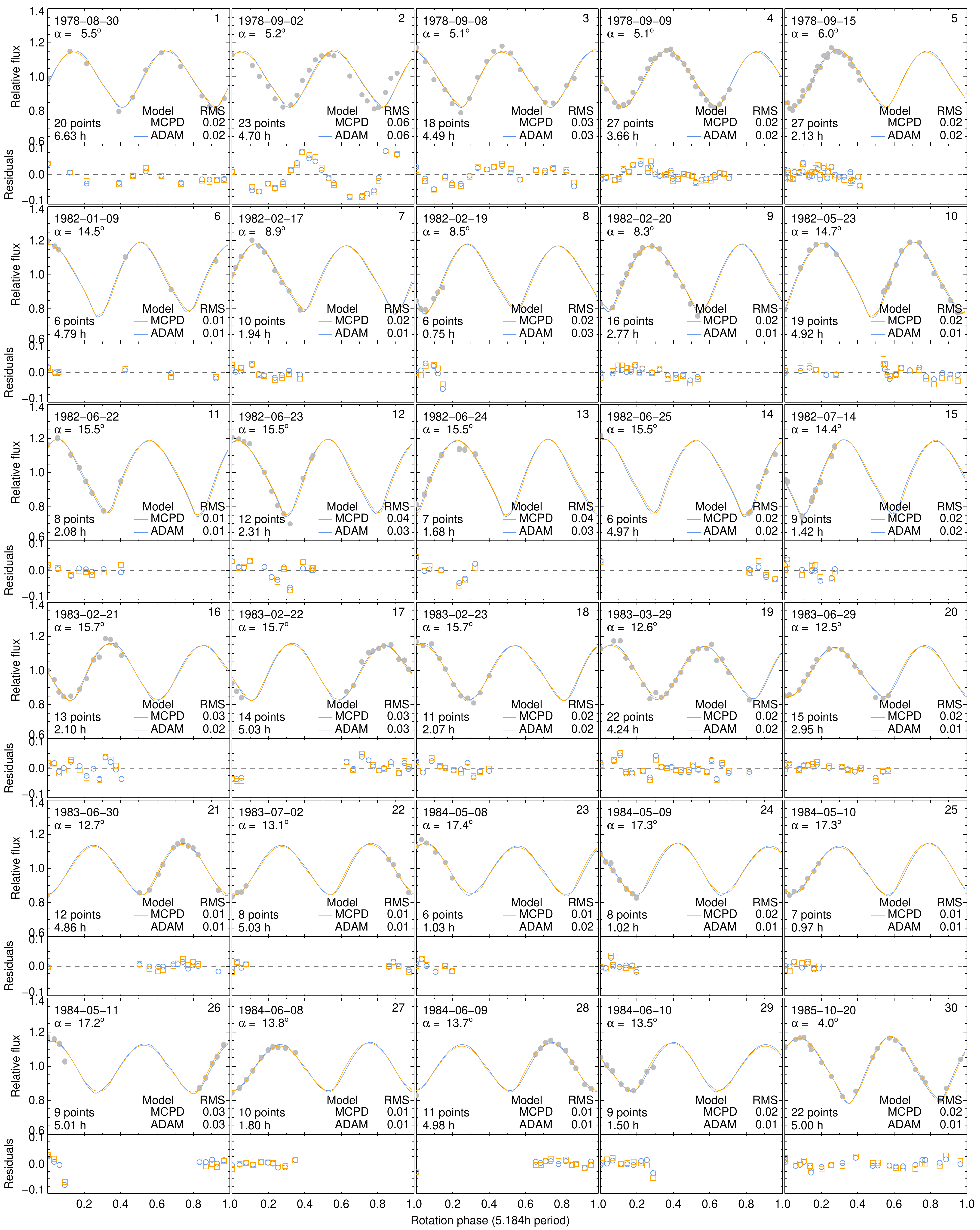}
  \end{subfigure}
  \caption[Optical lightcurves of Sylvia]{The optical lightcurves of Sylvia (grey dots),
    compared with the synthetic lightcurves generated with the \adam{} and \mpcd{} shape
    models (blue and orange lines). 
    On each panel, the observing date, number of points, duration
    of the lightcurve (in hours), and RMS residuals between the
    observations and the synthetic lightcurves from the shape model
    are displayed.
    In many cases, measurement uncertainties are not provided by the
    observers but can be estimated from the scatter of measurements. 
  \label{fig:app:lc}}
\end{figure*}
\clearpage
\begin{figure*}[h]
  \centering
  \ContinuedFloat
  \captionsetup{list=off}
  \begin{subfigure}{.9\textwidth}
    \includegraphics[width=\linewidth]{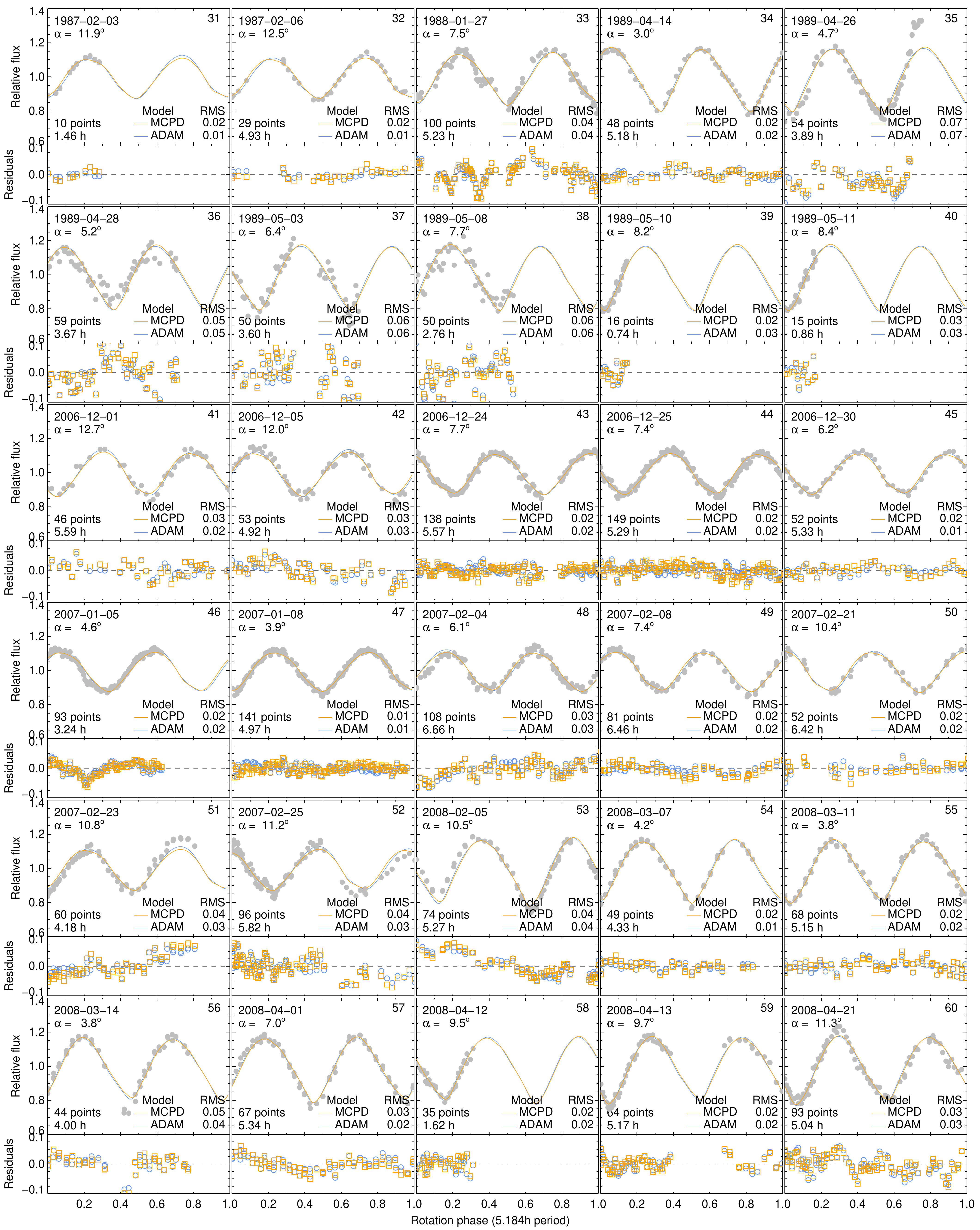}
  \end{subfigure}
  \caption{Suite of all lightcurve plots, as described in \Autoref{fig:app:lc}.}
\end{figure*}
\clearpage
\begin{figure*}[h]
  \centering
  \ContinuedFloat
  \captionsetup{list=off}
  \begin{subfigure}{.9\textwidth}
    \includegraphics[width=\linewidth]{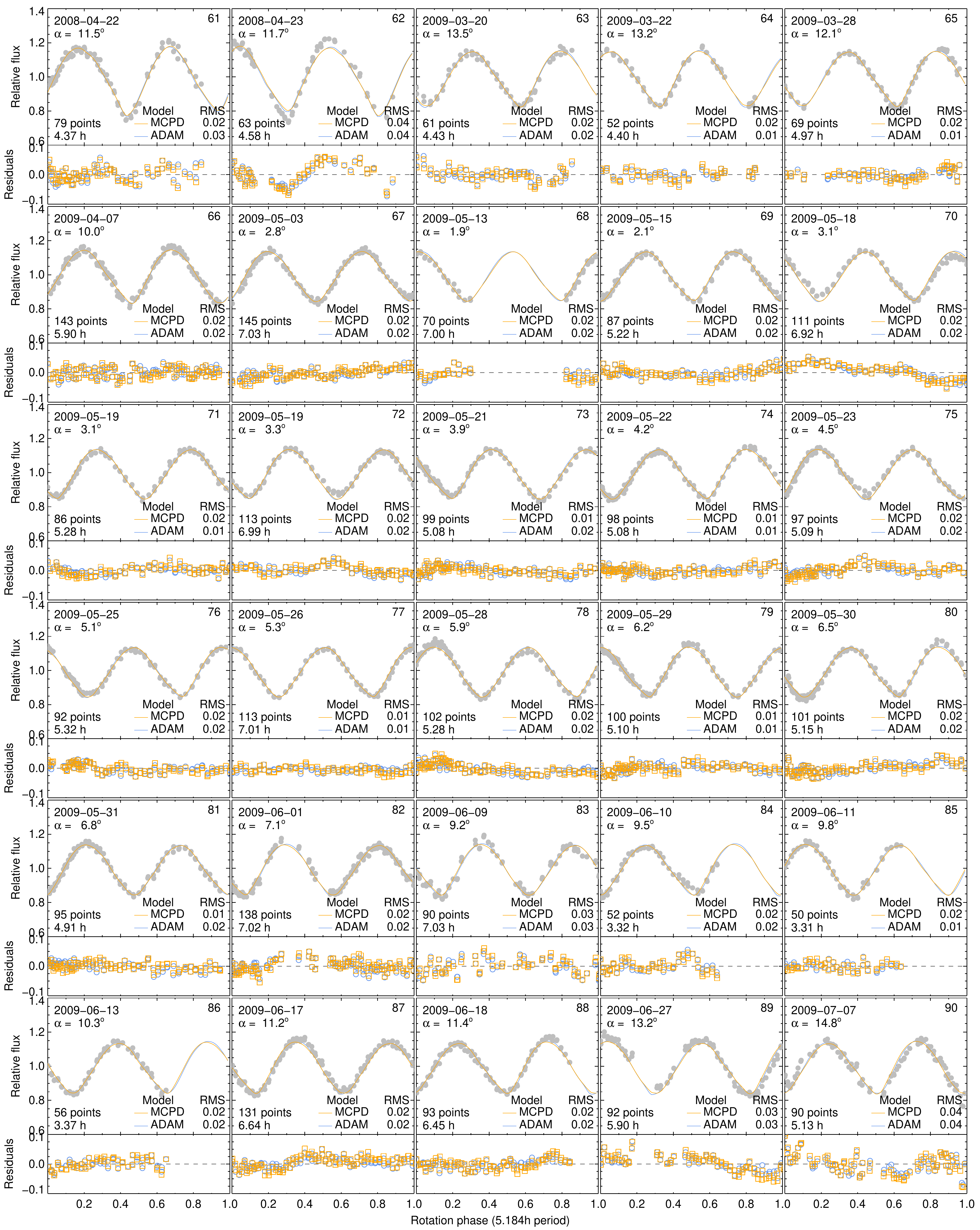}
  \end{subfigure}
  \caption{Suite continued from previous page.}
\end{figure*}
\clearpage
\begin{figure*}[h]
  \centering
  \ContinuedFloat
  \captionsetup{list=off}
  \begin{subfigure}{.9\textwidth}
    \includegraphics[width=\linewidth]{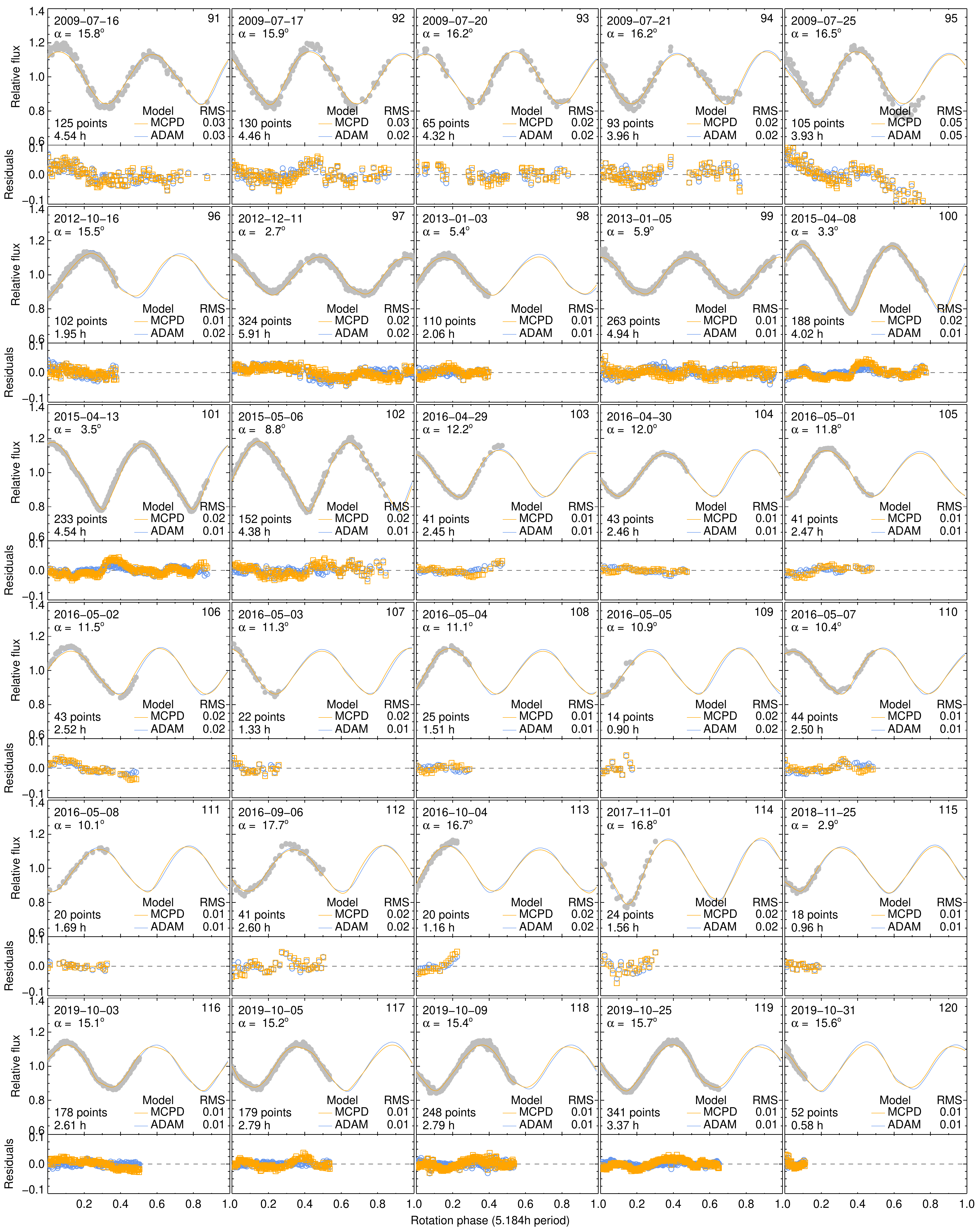}
  \end{subfigure}
  \caption{Suite continued from previous page.}
\end{figure*}

\onecolumn
\begin{center}
  \begin{longtable}{rcrrrlrcl}
    \caption{
      Date, duration ($\mathcal{L}$, in hours), number of points ($\mathcal{N}_p$), phase angle ($\alpha$), 
      filter, residual (against the shape model), IAU code, and observers, for each 
      lightcurve. \label{tab:lc}
    }\\

    \hline\hline
    & Date & \multicolumn{1}{c}{$\mathcal{L}$} & \multicolumn{1}{c}{$\mathcal{N}_p$} &
    \multicolumn{1}{c}{$\alpha$} & \multicolumn{1}{c}{Filter} & \multicolumn{1}{c}{RMS} &
    \multicolumn{1}{c}{IAU} & \multicolumn{1}{c}{Observers} \\
    && \multicolumn{1}{c}{(h)} && \multicolumn{1}{c}{(\degr)}&& \multicolumn{1}{c}{(mag)} \\
    \hline
    \endfirsthead

    \multicolumn{9}{c}{{\tablename\ \thetable{} -- continued from previous page}} \\ 
    \hline\hline
    & Date & \multicolumn{1}{c}{$\mathcal{L}$} & \multicolumn{1}{c}{$\mathcal{N}_p$} &
    \multicolumn{1}{c}{$\alpha$} & \multicolumn{1}{c}{Filter} & \multicolumn{1}{c}{RMS} &
    \multicolumn{1}{c}{IAU} & \multicolumn{1}{c}{Observers} \\
    && \multicolumn{1}{c}{(h)} && \multicolumn{1}{c}{(\degr)}&& \multicolumn{1}{c}{(mag)} \\
    \hline
    \endhead

    \hline \multicolumn{9}{r}{{Continued on next page}} \\ \hline
    \endfoot

    \hline
    \endlastfoot

      1 & 1978-08-30 &  6.6 &  20 &   5.5 & C     &  0.022 & 654 & \citet{1980Icar...43...20H}          \\
      2 & 1978-09-02 &  4.7 &  23 &   5.2 & C     &  0.060 & 654 & \citet{1980Icar...43...20H}          \\
      3 & 1978-09-08 &  4.5 &  18 &   5.1 & V     &  0.026 & 809 & \citet{1979AAS...38..269S} \\
      4 & 1978-09-09 &  3.7 &  27 &   5.1 & V     &  0.017 & 809 & \citet{1979AAS...38..269S} \\
      5 & 1978-09-15 &  2.1 &  27 &   6.0 & V     &  0.018 & 809 & \citet{1979AAS...38..269S} \\
      6 & 1982-01-09 &  4.8 &   6 &  14.5 & V     &  0.012 & 695 & \citet{1987Icar...70..191W}     \\
      7 & 1982-02-17 &  1.9 &  10 &   8.9 & V     &  0.012 & 695 & \citet{1987Icar...70..191W}     \\
      8 & 1982-02-19 &  0.8 &   6 &   8.5 & V     &  0.029 & 695 & \citet{1987Icar...70..191W}     \\
      9 & 1982-02-20 &  2.8 &  16 &   8.3 & V     &  0.013 & 695 & \citet{1987Icar...70..191W}     \\
     10 & 1982-05-23 &  4.9 &  19 &  14.7 & V     &  0.015 & 695 & \citet{1987Icar...70..191W}     \\
     11 & 1982-06-22 &  2.1 &   8 &  15.5 & V     &  0.009 & 695 & \citet{1987Icar...70..191W}     \\
     12 & 1982-06-23 &  2.3 &  12 &  15.5 & V     &  0.032 & 695 & \citet{1987Icar...70..191W}     \\
     13 & 1982-06-24 &  1.7 &   7 &  15.5 & V     &  0.032 & 695 & \citet{1987Icar...70..191W}     \\
     14 & 1982-06-25 &  5.0 &   6 &  15.5 & V     &  0.019 & 695 & \citet{1987Icar...70..191W}     \\
     15 & 1982-07-14 &  1.4 &   9 &  14.4 & V     &  0.021 & 695 & \citet{1987Icar...70..191W}     \\
     16 & 1983-02-21 &  2.1 &  13 &  15.7 & V     &  0.023 & 695 & \citet{1987Icar...70..191W}     \\
     17 & 1983-02-22 &  5.0 &  14 &  15.7 & V     &  0.027 & 695 & \citet{1987Icar...70..191W}     \\
     18 & 1983-02-23 &  2.1 &  11 &  15.7 & V     &  0.018 & 695 & \citet{1987Icar...70..191W}     \\
     19 & 1983-03-29 &  4.2 &  22 &  12.6 & V     &  0.023 & 695 & \citet{1987Icar...70..191W}     \\
     20 & 1983-06-29 &  2.9 &  15 &  12.5 & V     &  0.014 & 695 & \citet{1987Icar...70..191W}     \\
     21 & 1983-06-30 &  4.9 &  12 &  12.7 & V     &  0.009 & 695 & \citet{1987Icar...70..191W}     \\
     22 & 1983-07-02 &  5.0 &   8 &  13.1 & V     &  0.009 & 695 & \citet{1987Icar...70..191W}     \\
     23 & 1984-05-08 &  1.0 &   6 &  17.4 & V     &  0.017 & 695 & \citet{1987Icar...70..191W}     \\
     24 & 1984-05-09 &  1.0 &   8 &  17.3 & V     &  0.013 & 695 & \citet{1987Icar...70..191W}     \\
     25 & 1984-05-10 &  1.0 &   7 &  17.3 & V     &  0.009 & 695 & \citet{1987Icar...70..191W}     \\
     26 & 1984-05-11 &  5.0 &   9 &  17.2 & V     &  0.028 & 695 & \citet{1987Icar...70..191W}     \\
     27 & 1984-06-08 &  1.8 &  10 &  13.8 & V     &  0.008 & 695 & \citet{1987Icar...70..191W}     \\
     28 & 1984-06-09 &  5.0 &  11 &  13.7 & V     &  0.014 & 695 & \citet{1987Icar...70..191W}     \\
     29 & 1984-06-10 &  1.5 &   9 &  13.5 & V     &  0.013 & 695 & \citet{1987Icar...70..191W}     \\
     30 & 1985-10-20 &  5.0 &  22 &   4.0 & V     &  0.013 & 695 & \citet{1987Icar...70..191W}     \\
     31 & 1987-02-03 &  1.5 &  10 &  11.9 & V     &  0.006 & 695 & \citet{1990Icar...86..402W}     \\
     32 & 1987-02-06 &  4.9 &  29 &  12.5 & V     &  0.012 & 695 & \citet{1990Icar...86..402W}     \\
     33 & 1988-01-27 &  5.2 & 100 &   7.5 & --    &  0.037 &  156 & \citet{1989MmSAI..60..195B} \\
     34 & 1989-04-14 &  5.2 &  48 &   3.0 & V     &  0.015 & 695 & \citet{1990Icar...86..402W}     \\
     35 & 1989-04-26 &  3.9 &  54 &   4.7 & B     &  0.068 &  -- & \citet{1992ATsir1552...27P}  \\
     36 & 1989-04-28 &  3.7 &  59 &   5.2 & B     &  0.054 &  -- & \citet{1992ATsir1552...27P}  \\
     37 & 1989-05-03 &  3.6 &  50 &   6.4 & R     &  0.061 &  -- & \citet{1992ATsir1552...27P}  \\
     38 & 1989-05-08 &  2.8 &  50 &   7.7 & B     &  0.064 &  -- & \citet{1992ATsir1552...27P}  \\
     39 & 1989-05-10 &  0.7 &  16 &   8.2 & R     &  0.025 &  -- & \citet{1992ATsir1552...27P}  \\
     40 & 1989-05-11 &  0.9 &  15 &   8.4 & R     &  0.029 &  -- & \citet{1992ATsir1552...27P}  \\
     41 & 2006-11-30 &  5.6 &  46 &  12.7 & clear &  0.024 & 950 & SuperWASP - J. Grice                      \\
     42 & 2006-12-04 &  4.9 &  53 &  12.0 & clear &  0.031 & 950 & SuperWASP - J. Grice                      \\
     43 & 2006-12-24 &  5.6 & 138 &   7.7 & C     &  0.015 & D19 & \rev{\citet{2016AA...586A.108H}} \\
     44 & 2006-12-25 &  5.3 & 149 &   7.4 & C     &  0.018 & D19 & \rev{\citet{2016AA...586A.108H}} \\
     45 & 2006-12-29 &  5.3 &  52 &   6.2 & clear &  0.013 & 950 & SuperWASP - J. Grice                      \\
     46 & 2007-01-05 &  3.2 &  93 &   4.6 & C     &  0.021 & D19 & \rev{\citet{2016AA...586A.108H}} \\
     47 & 2007-01-08 &  5.0 & 141 &   3.9 & C     &  0.013 & D19 & \rev{\citet{2016AA...586A.108H}} \\
     48 & 2007-02-03 &  6.7 & 108 &   6.1 & clear &  0.029 & 950 & SuperWASP - J. Grice                      \\
     49 & 2007-02-08 &  6.5 &  81 &   7.4 & clear &  0.018 & 950 & SuperWASP - J. Grice                      \\
     50 & 2007-02-21 &  6.4 &  52 &  10.4 & clear &  0.017 & 950 & SuperWASP - J. Grice                      \\
     51 & 2007-02-23 &  4.2 &  60 &  10.8 & clear &  0.029 & 950 & SuperWASP - J. Grice                      \\
     52 & 2007-02-25 &  5.8 &  96 &  11.2 & clear &  0.032 & 950 & SuperWASP - J. Grice                      \\
     53 & 2008-02-04 &  5.3 &  74 &  10.5 & clear &  0.040 & 950 & SuperWASP - J. Grice                      \\
     54 & 2008-03-06 &  4.3 &  49 &   4.2 & clear &  0.014 & 950 & SuperWASP - J. Grice                      \\
     55 & 2008-03-10 &  5.1 &  68 &   3.8 & clear &  0.020 & 950 & SuperWASP - J. Grice                      \\
     56 & 2008-03-13 &  4.0 &  44 &   3.8 & clear &  0.040 & 950 & SuperWASP - J. Grice                      \\
     57 & 2008-04-01 &  5.3 &  67 &   7.0 & clear &  0.024 & 950 & SuperWASP - J. Grice                      \\
     58 & 2008-04-12 &  1.6 &  35 &   9.5 & clear &  0.021 & 950 & SuperWASP - J. Grice                      \\
     59 & 2008-04-13 &  5.2 &  64 &   9.7 & clear &  0.024 & 950 & SuperWASP - J. Grice                      \\
     60 & 2008-04-21 &  5.0 &  93 &  11.3 & clear &  0.033 & 950 & SuperWASP - J. Grice                      \\
     61 & 2008-04-22 &  4.4 &  79 &  11.5 & clear &  0.025 & 950 & SuperWASP - J. Grice                      \\
     62 & 2008-04-23 &  4.6 &  63 &  11.7 & clear &  0.040 & 950 & SuperWASP - J. Grice                      \\
     63 & 2009-03-19 &  4.4 &  61 &  13.5 & clear &  0.023 & 950 & SuperWASP - J. Grice                      \\
     64 & 2009-03-21 &  4.4 &  52 &  13.2 & clear &  0.015 & 950 & SuperWASP - J. Grice                      \\
     65 & 2009-03-27 &  5.0 &  69 &  12.1 & clear &  0.014 & 950 & SuperWASP - J. Grice                      \\
     66 & 2009-04-06 &  5.9 & 143 &  10.0 & clear &  0.019 & 950 & SuperWASP - J. Grice                      \\
     67 & 2009-05-03 &  7.0 & 145 &   2.8 & clear &  0.018 & 950 & SuperWASP - J. Grice                      \\
     68 & 2009-05-13 &  7.0 &  70 &   1.9 & clear &  0.021 & 950 & SuperWASP - J. Grice                      \\
     69 & 2009-05-14 &  5.2 &  87 &   2.1 & clear &  0.017 & 950 & SuperWASP - J. Grice                      \\
     70 & 2009-05-18 &  6.9 & 111 &   3.1 & clear &  0.024 & 950 & SuperWASP - J. Grice                      \\
     71 & 2009-05-18 &  5.3 &  86 &   3.1 & clear &  0.014 & 950 & SuperWASP - J. Grice                      \\
     72 & 2009-05-19 &  7.0 & 113 &   3.3 & clear &  0.017 & 950 & SuperWASP - J. Grice                      \\
     73 & 2009-05-21 &  5.1 &  99 &   3.9 & clear &  0.017 & 950 & SuperWASP - J. Grice                      \\
     74 & 2009-05-22 &  5.1 &  98 &   4.2 & clear &  0.014 & 950 & SuperWASP - J. Grice                      \\
     75 & 2009-05-23 &  5.1 &  97 &   4.5 & clear &  0.017 & 950 & SuperWASP - J. Grice                      \\
     76 & 2009-05-25 &  5.3 &  92 &   5.1 & clear &  0.016 & 950 & SuperWASP - J. Grice                      \\
     77 & 2009-05-26 &  7.0 & 113 &   5.3 & clear &  0.013 & 950 & SuperWASP - J. Grice                      \\
     78 & 2009-05-28 &  5.3 & 102 &   5.9 & clear &  0.017 & 950 & SuperWASP - J. Grice                      \\
     79 & 2009-05-29 &  5.1 & 100 &   6.2 & clear &  0.015 & 950 & SuperWASP - J. Grice                      \\
     80 & 2009-05-30 &  5.1 & 101 &   6.5 & clear &  0.018 & 950 & SuperWASP - J. Grice                      \\
     81 & 2009-05-31 &  4.9 &  95 &   6.8 & clear &  0.015 & 950 & SuperWASP - J. Grice                      \\
     82 & 2009-06-01 &  7.0 & 138 &   7.1 & clear &  0.019 & 950 & SuperWASP - J. Grice                      \\
     83 & 2009-06-09 &  7.0 &  90 &   9.2 & clear &  0.027 & 950 & SuperWASP - J. Grice                      \\
     84 & 2009-06-10 &  3.3 &  52 &   9.5 & clear &  0.022 & 950 & SuperWASP - J. Grice                      \\
     85 & 2009-06-11 &  3.3 &  50 &   9.8 & clear &  0.014 & 950 & SuperWASP - J. Grice                      \\
     86 & 2009-06-13 &  3.4 &  56 &  10.3 & clear &  0.016 & 950 & SuperWASP - J. Grice                      \\
     87 & 2009-06-17 &  6.6 & 131 &  11.2 & clear &  0.020 & 950 & SuperWASP - J. Grice                      \\
     88 & 2009-06-18 &  6.4 &  93 &  11.4 & clear &  0.015 & 950 & SuperWASP - J. Grice                      \\
     89 & 2009-06-27 &  5.9 &  92 &  13.2 & clear &  0.027 & 950 & SuperWASP - J. Grice                      \\
     90 & 2009-07-07 &  5.1 &  90 &  14.8 & clear &  0.036 & 950 & SuperWASP - J. Grice                      \\
     91 & 2009-07-16 &  4.5 & 125 &  15.8 & clear &  0.026 & 950 & SuperWASP - J. Grice                      \\
     92 & 2009-07-17 &  4.5 & 130 &  15.9 & clear &  0.021 & 950 & SuperWASP - J. Grice                      \\
     93 & 2009-07-20 &  4.3 &  65 &  16.2 & clear &  0.017 & 950 & SuperWASP - J. Grice                      \\
     94 & 2009-07-21 &  4.0 &  93 &  16.2 & clear &  0.023 & 950 & SuperWASP - J. Grice                      \\
     95 & 2009-07-25 &  3.9 & 105 &  16.5 & clear &  0.047 & 950 & SuperWASP - J. Grice                      \\
     96 & 2012-10-16 &  2.0 & 102 &  15.5 & clear &  0.018 & Z74 & CdR: F. Soldan \\
     97 & 2012-12-11 &  5.9 & 324 &   2.7 & clear &  0.019 & 511 & CdR: J. Strajnic \\
     98 & 2013-01-03 &  2.1 & 110 &   5.4 & R     &  0.008 & 627 & CdR: R. Roy \\
     99 & 2013-01-05 &  4.9 & 263 &   5.9 & R     &  0.013 & 627 & CdR: R. Roy \\
    100 & 2015-04-08 &  4.0 & 188 &   3.3 & clear &  0.011 & A12 & CdR: F. Manzini \\
    101 & 2015-04-13 &  4.5 & 233 &   3.5 & clear &  0.010 & A12 & CdR: F. Manzini \\
    102 & 2015-05-07 &  4.4 & 152 &   8.8 & clear &  0.012 & Z74 & CdR: D. Romeuf \\
    103 & 2016-04-29 &  2.4 &  41 &  12.2 & R     &  0.010 &  181 & Vachier, Klotz, Teng, Peyrot, Thierry, Berthier \\
    104 & 2016-04-30 &  2.5 &  43 &  12.0 & R     &  0.009 &  181 & Vachier, Klotz, Teng, Peyrot, Thierry, Berthier \\
    105 & 2016-05-01 &  2.5 &  41 &  11.8 & R     &  0.010 &  181 & Vachier, Klotz, Teng, Peyrot, Thierry, Berthier \\
    106 & 2016-05-02 &  2.5 &  43 &  11.5 & R     &  0.017 &  181 & Vachier, Klotz, Teng, Peyrot, Thierry, Berthier \\
    107 & 2016-05-03 &  1.3 &  22 &  11.3 & R     &  0.013 &  181 & Vachier, Klotz, Teng, Peyrot, Thierry, Berthier \\
    108 & 2016-05-04 &  1.5 &  25 &  11.1 & R     &  0.008 &  181 & Vachier, Klotz, Teng, Peyrot, Thierry, Berthier \\
    109 & 2016-05-05 &  0.9 &  14 &  10.9 & R     &  0.015 &  181 & Vachier, Klotz, Teng, Peyrot, Thierry, Berthier \\
    110 & 2016-05-07 &  2.5 &  44 &  10.4 & R     &  0.014 &  181 & Vachier, Klotz, Teng, Peyrot, Thierry, Berthier \\
    111 & 2016-05-08 &  1.7 &  20 &  10.1 & R     &  0.009 &  181 & Vachier, Klotz, Teng, Peyrot, Thierry, Berthier \\
    112 & 2016-09-06 &  2.6 &  41 &  17.7 & R     &  0.017 &  181 & Vachier, Klotz, Teng, Peyrot, Thierry, Berthier \\
    113 & 2016-10-04 &  1.2 &  20 &  16.7 & R     &  0.015 &  181 & Vachier, Klotz, Teng, Peyrot, Thierry, Berthier \\
    114 & 2017-11-01 &  1.6 &  24 &  16.8 & R     &  0.023 &  181 & Vachier, Klotz, Teng, Peyrot, Thierry, Berthier \\
    115 & 2018-11-25 &  1.0 &  18 &   2.9 & R     &  0.007 &  181 & Vachier, Klotz, Teng, Peyrot, Thierry, Berthier \\
    116 & 2019-10-03 &  2.6 & 178 &  15.1 & Rc    &  0.006 &  Z53 & M. Ferrais and E. Jehin \\
    117 & 2019-10-05 &  2.8 & 179 &  15.2 & Rc    &  0.006 &  Z53 & M. Ferrais and E. Jehin \\
    118 & 2019-10-09 &  2.8 & 248 &  15.4 & Rc    &  0.009 &  Z53 & M. Ferrais and E. Jehin \\
    119 & 2019-10-25 &  3.4 & 341 &  15.7 & Rc    &  0.006 &  I40 & M. Ferrais and E. Jehin \\
    120 & 2019-10-31 &  0.6 &  52 &  15.6 & Rc    &  0.009 &  I40 & M. Ferrais and E. Jehin \\
    121 & 2019-11-08 &  1.6 & 140 &  15.2 & Rc    &  0.008 &  Z53 & M. Ferrais and E. Jehin \\
\hline
  \end{longtable}
\end{center}
\twocolumn

\section{Stellar occultations\label{sec:app:occ}}

\begin{figure*}[ht]
    \centering
    \includegraphics[width=\hsize]{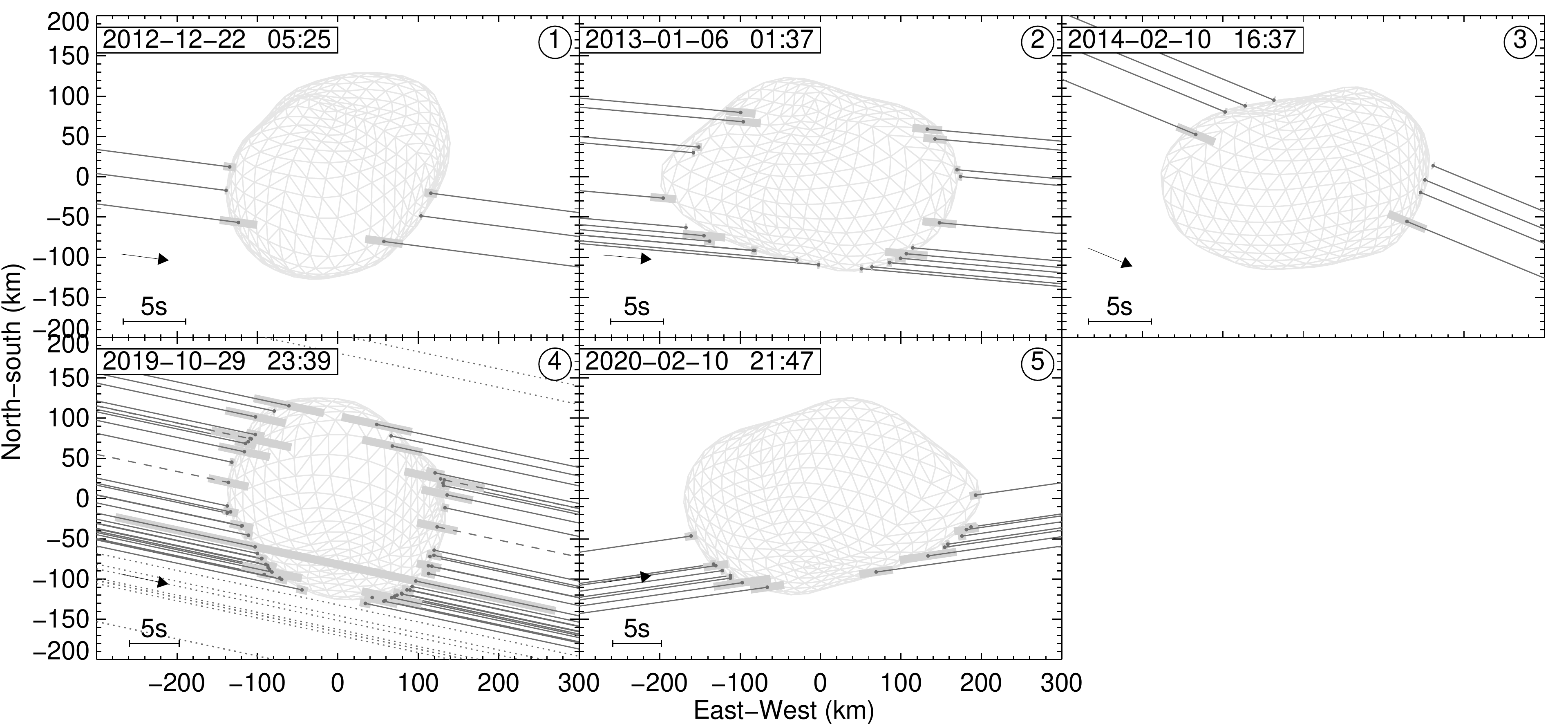}
    \caption{The five stellar occultations by Sylvia used in this work,
    compared with the shape model projected on the 
    plane of the sky for the times of the
    occultations.
    \rev{The dark grey lines represent the chords and the thick light grey lines their uncertainties.}}
    \label{fig:app:occ}
\end{figure*}

\onecolumn
\begin{longtable}{l}
\caption{\label{tab:occ}List of observers of the five stellar occultations.}\\
\hline
 Observer \\ \hline\hline

\endfirsthead
\caption{continued.}\\

\hline
 Observer \\ \hline\hline
\endhead
\hline
\endfoot

\multicolumn{1}{c} {\textbf{(87) Sylvia 2012-12-22}} \\
T. Campbell, Tampa, FL \\
R. Venable, Ft. Drum,  FL \\
T. Campbell, Sarasota, FL \\
R. Venable, Weston, FL \\
R. Venable, Pennsuco, FL \\

\multicolumn{1}{c} {\textbf{(87) Sylvia 2013-01-06}} \\
Francois Colas, FR    \\
Vasilis Metallinos, GR                \\
Hilari Pallares, ES                   \\
Ricard Casas, ES                      \\
C. Perello, A. Selva, ES               \\
Joan Lopez, ES                        \\
Dominique Albanese, FR                \\
L. Brunetto, J-M. Mari, A. Lopez, C. Bouteiller, FR    \\
Raymond Poncy, FR                     \\
J. Lecacheux, O. Lecacheux, FR           \\
E. Frappa, A. Klotz, FR                \\
M Devogele, P. Bendjoya, L. Abe, O. Suarez, J.-P. Rivet, FR    \\
Daniel Verilhac, FR                   \\
Pierre Dubreuil, FR                   \\
Paolo Tanga, FR                       \\
Guy Brabant, FR                       \\
Eric Frappa, FR                       \\
Luc Arnold, FR                        \\
E. Frappa, M. Lavayssiere, FR            \\
J. Lecacheux, S. Moindrot, FR            \\
Laurent Bernasconi, FR                \\
Claude Peguet, FR                     \\
Jean-Louis Penninckx, FR              \\
Marc Bretton, FR                      \\
Alain Figer, FR                       \\
Vincent Fristot, FR                   \\
S. Sposetti, A. Manna, IT              \\
U. Quadri, L. Strabla, R. Girelli, A. Quadri, IT     \\
Simone Bolzoni, IT                    \\
Albino Carbognani, IT                 \\
Carlo Gualdoni, IT                    \\
Stefano Sposetti, CH                  \\

\multicolumn{1}{c} {\textbf{(87) Sylvia 2014-02-10}} \\
H. Watanabe, Inabe, Mie, Japan         \\
H. Watanabe, Tarui, Gifu, Japan        \\
M. Ishida, Joyo, Kyoto, Japan          \\
M. Ida, Higashiomi, Shiga, Japan       \\
T. Terada, Yaotsu, Gifu, Japan         \\

\multicolumn{1}{c} {\textbf{(87) Sylvia 2019-10-29}} \\
Alex Pratt, UK \\ 
O. Canales et al., ES\\ 
P. Andre \rev{et al., FR}\\ 
P. Andre \rev{et al., FR}\\ 
Carles Schnabel, ES \\ 
Eric Frappa, FR \\ 
C. Perello/A., Selva \\ 
Jean-Marc Lechopier, FR \\ 
F. Van, Den \\ 
A. Martin/E., Arbouch \\ 
O. Schreurs, et \\ 
A. Malvache, et \\ 
A. Wuensche, et \\ 
Patrick Sogorb, FR \\ 
Matthieu Conjat, FR \\ 
Friedhelm Huebner, DE \\ 
Oliver Kloes, DE \\ 
J. Ohlert, et \\ 
Michael Koch, DE \\ 
Bernd Gaehrken, DE \\ 
Thomas Berthold, DE \\ 
Ralf Koehrbrueck, DE \\ 
Ralf Koehrbrueck, DE \\ 
Jiri Polak, CZ \\ 
Libor Smid, CZ \\ 
Christian Weber, DE \\ 
Sirko Molau, DE \\ 
Karel Halir, CZ \\ 
Nikolai Wuensche, DE \\ 
Nikolai Wuensche, DE \\ 
Dieter Ewald, DE \\ 
Peter Enskonatus, DE \\ 
Jiri Kubanek, CZ \\ 
Jiri Lev, CZ \\ 
Michal Rottenborn, CZ \\ 
Jan Zahajsky, CZ \\ 
Jaromir Jindra, CZ \\ 
Stanislav Holy, CZ \\ 
Jan Manek, CZ \\ 
Ladislav Cervinka, CZ \\ 
Martin Gembec, CZ \\ 
Milan Antos, CZ \\ 
Tomasz Kluwak, PL \\ 
Krzysztof Kaminski, PL \\ 
Anna Marciniak, PL \\ 
A. Wrembel \rev{et al., PL}\\ 
Marian Urbanik, SK \\ 
Marek Zawilski, PL \\ 
Dariusz Miller, PL \\ 
Donatas Tamonis, LT \\ 

\multicolumn{1}{c} {\textbf{(87) Sylvia 2020-02-10}} \\
Peter Birtwhistle, UK \\
John Talbot, UK \\
Peter Tickner, UK \\
Adrian Jones, UK \\
Malcolm Jennings, UK \\
Simon Kidd, UK \\
Philip Denyer, UK \\
Petr Zeleny, CZ \\

\hline
 \end{longtable}
\twocolumn

\section{Astrometry of Romulus and Remus\label{sec:app:sat}}

\onecolumn
\begin{center}
  \begin{longtable}{cclllrrrrrrr}
  \caption[Astrometry of Romulus]{Astrometry of Sylvia's satellite Romulus.
    Date, mid-observing time (UTC), telescope, camera, filter, 
    astrometry ($X$ is aligned with Right Ascension, and $Y$ with Declination, and 
    $o$ and $c$ indices stand for observed and computed positions),
    uncertainty ($\sigma$), 
    and photometry (magnitude difference $\Delta M$ with uncertainty $\delta M$).
    \label{tab:sat1}
  }\\

    \hline\hline
     Date & UTC & Tel. & Cam. & Filter &
     \multicolumn{1}{c}{$X_o$} &
     \multicolumn{1}{c}{$Y_o$} &
     \multicolumn{1}{c}{$X_{o-c}$} &
     \multicolumn{1}{c}{$Y_{o-c}$} &
     \multicolumn{1}{c}{$\sigma$} &
     \multicolumn{1}{c}{$\Delta M$} &
     \multicolumn{1}{c}{$\delta M$} \\
    &&&&& 
     \multicolumn{1}{c}{(mas)} & \multicolumn{1}{c}{(mas)} & 
     \multicolumn{1}{c}{(mas)} & \multicolumn{1}{c}{(mas)} & 
     \multicolumn{1}{c}{(mas)} & 
     \multicolumn{1}{c}{(mag)} & \multicolumn{1}{c}{(mag)}  \\ 
    \hline
    \endfirsthead

    \multicolumn{11}{c}{{\tablename\ \thetable{} -- continued from previous page}} \\ 
    \hline\hline
     Date & UTC & Tel. & Cam. & Filter &
     \multicolumn{1}{c}{$X_o$} &
     \multicolumn{1}{c}{$Y_o$} &
     \multicolumn{1}{c}{$X_{o-c}$} &
     \multicolumn{1}{c}{$Y_{o-c}$} &
     \multicolumn{1}{c}{$\sigma$} &
     \multicolumn{1}{c}{$\Delta M$} &
     \multicolumn{1}{c}{$\delta M$} \\
    &&&&& 
     \multicolumn{1}{c}{(mas)} & \multicolumn{1}{c}{(mas)} & 
     \multicolumn{1}{c}{(mas)} & \multicolumn{1}{c}{(mas)} & 
     \multicolumn{1}{c}{(mas)} & 
     \multicolumn{1}{c}{(mag)} & \multicolumn{1}{c}{(mag)}  \\ 
    \hline
    \endhead

    \hline \multicolumn{11}{r}{{Continued on next page}} \\ \hline
    \endfoot

    \hline
    \endlastfoot

2002-05-07 & 09:37:02.42 & Keck & NIRC2 & H & -595.9 & -1.0 & -1.7 & 5.7 & 10.0 & 6.8 & 0.2  \\ 
2002-05-08 & 09:42:04.29 & Keck & NIRC2 & H & -208.4 & 52.8 & -7.4 & 0.2 & 10.0 &     &      \\ 
2003-06-05 & 12:54:41.82 & Keck & NIRC2 & Ks & -763.0 & 75.1 & -3.2 & 9.0 & 10.0 & 6.5 & 0.0  \\ 
2003-08-14 & 07:02:32.56 & Keck & NIRC2 & H & -305.3 & 336.7 & 3.9 & 9.3 & 10.0 & 6.5 & 0.2  \\ 
2003-08-14 & 07:10:03.39 & Keck & NIRC2 & H & -302.9 & 333.9 & 1.0 & 5.7 & 10.0 & 6.6 & 0.3  \\ 
2004-07-25 & 10:22:15.02 & VLT & NACO & H & -763.3 & -8.9 & -13.7 & -3.5 & 13.3 & 6.0 & 0.2  \\ 
2004-08-10 & 07:17:15.21 & VLT & NACO & Ks & 374.3 & 142.5 & -7.0 & -5.0 & 13.3 & 6.2 & 0.1  \\ 
2004-08-28 & 06:31:54.59 & VLT & NACO & Ks & 105.8 & 212.3 & -1.3 & 4.7 & 13.3 & 8.4 & 1.2  \\ 
2004-08-29 & 07:18:17.00 & VLT & NACO & Ks & 775.5 & -91.2 & -3.9 & 16.5 & 13.3 & 6.5 & 0.1  \\ 
2004-08-29 & 07:26:16.37 & VLT & NACO & H & 782.8 & -93.6 & 6.2 & 15.9 & 13.3 & 7.2 & 1.0  \\ 
2004-08-29 & 08:46:06.90 & VLT & NACO & Ks & 751.8 & -116.5 & 6.1 & 10.7 & 13.3 & 6.6 & 0.2  \\ 
2004-09-01 & 05:54:49.70 & VLT & NACO & Ks & 544.1 & 136.8 & 0.1 & -3.4 & 13.3 & 5.8 &      \\ 
2004-09-01 & 06:03:51.58 & VLT & NACO & H & 554.0 & 136.8 & 3.2 & -1.6 & 13.3 & 5.9 & 0.0  \\ 
2004-09-01 & 08:25:28.47 & VLT & NACO & Ks & 652.3 & 109.2 & 4.8 & 2.3 & 13.3 & 5.9 & 0.0  \\ 
2004-09-03 & 07:08:59.76 & VLT & NACO & Ks & -760.6 & -58.3 & -15.0 & 2.9 & 13.3 & 5.8 & 0.1  \\ 
2004-09-04 & 08:41:05.05 & VLT & NACO & Ks & -146.0 & 230.4 & 2.2 & 4.2 & 13.3 & 6.1 & 1.5  \\ 
2004-09-05 & 04:09:23.01 & VLT & NACO & Ks & 786.3 & 46.8 & 9.1 & 7.4 & 13.3 & 6.2 & 0.1  \\ 
2004-09-05 & 08:15:16.29 & VLT & NACO & Ks & 831.8 & -18.6 & 6.2 & 8.9 & 13.3 & 5.8 & 0.1  \\ 
2004-09-06 & 03:49:57.98 & VLT & NACO & Ks & 174.7 & -223.2 & -0.5 & 4.7 & 13.3 & 6.8 & 0.2  \\ 
2004-09-07 & 02:36:18.09 & VLT & NACO & Ks & -832.7 & 3.6 & -15.5 & -3.5 & 13.3 & 5.9 & 0.1  \\ 
2004-09-07 & 09:24:50.66 & VLT & NACO & Ks & -787.8 & 103.2 & -11.4 & -11.0 & 13.3 & 5.9 & 0.2  \\ 
2004-10-19 & 00:27:03.03 & VLT & NACO & Ks & 751.6 & -37.6 & 20.6 & 8.0 & 13.3 & 6.8 & 0.1  \\ 
2004-10-20 & 00:08:21.61 & VLT & NACO & Ks & -88.0 & -210.2 & 16.5 & -14.7 & 13.3 & 6.3 & 1.2  \\ 
2004-11-02 & 07:34:31.11 & Gemini & NIRI & Kp & 631.9 & 43.1 & 16.3 & 1.0 & 21.9 & 5.7 & 0.0  \\ 
2005-10-18 & 12:33:05.50 & Gemini & NIRI & Ks & 569.4 & -109.7 & 5.9 & -9.5 & 21.9 & 6.5 & 1.2  \\ 
2005-11-01 & 13:13:18.10 & Gemini & NIRI & Ks & 2.8 & -259.7 & 6.8 & -5.3 & 21.9 & 6.0 & 1.2  \\ 
2005-11-06 & 08:01:12.70 & Gemini & NIRI & Ks & 709.8 & 166.0 & 14.0 & -1.1 & 21.9 & 6.1 & 0.2  \\ 
2005-11-06 & 08:31:28.38 & Gemini & NIRI & Ks & 694.3 & 173.6 & 10.0 & -0.9 & 21.9 & 5.8 & 0.2  \\ 
2005-11-06 & 08:36:45.45 & Gemini & NIRI & Ks & 693.2 & 178.4 & 11.0 & 2.6 & 21.9 & 5.8 & 0.1  \\ 
2005-12-20 & 10:01:05.00 & Gemini & NIRI & Ks & 250.1 & 256.7 & 4.5 & 3.7 & 21.9 & 6.2 & 0.3  \\ 
2005-12-20 & 10:08:25.01 & Gemini & NIRI & Ks & 244.4 & 265.7 & 4.7 & 12.5 & 21.9 & 5.9 & 2.0  \\ 
2005-12-21 & 08:42:04.55 & Gemini & NIRI & Ks & -689.8 & 22.4 & -14.2 & 5.1 & 21.9 & 5.9 & 0.0  \\ 
2005-12-21 & 08:47:19.69 & Gemini & NIRI & Ks & -697.8 & 4.1 & -20.9 & -11.6 & 21.9 & 6.3 & 1.2  \\ 
2006-01-01 & 10:33:41.50 & Gemini & NIRI & Ks & -675.8 & -50.3 & -2.6 & -0.5 & 21.9 & 5.7 & 0.6  \\ 
2006-01-01 & 10:43:05.31 & Gemini & NIRI & Ks & -677.2 & -52.3 & -3.9 & 0.2 & 21.9 & 5.5 & 0.1  \\ 
2006-01-06 & 09:14:34.20 & Gemini & NIRI & Ks & 423.6 & -156.7 & 15.3 & -2.0 & 21.9 & 6.0 & 0.1  \\ 
2006-01-06 & 09:20:57.01 & Gemini & NIRI & Ks & 420.7 & -153.5 & 8.4 & -0.2 & 21.9 & 6.0 & 0.0  \\ 
2006-12-12 & 16:10:57.12 & Keck & NIRC2 & Kp & 546.9 & 194.1 & 10.0 & -7.5 & 10.0 & 7.5 & 0.1  \\ 
2008-01-21 & 11:01:10.12 & Keck & NIRC2 & Kp & 475.6 & 14.5 & 1.4 & -20.1 & 10.0 & 6.0 & 3.3  \\ 
2008-01-21 & 11:11:07.59 & Keck & NIRC2 & Kp & 474.7 & 33.3 & 4.6 & -1.8 & 10.0 & 6.5 & 0.2  \\ 
2008-01-21 & 11:24:55.26 & Keck & NIRC2 & Kp & 464.6 & 31.0 & 0.3 & -4.7 & 10.0 & 6.5 & 0.1  \\ 
2008-01-21 & 11:56:28.79 & Keck & NIRC2 & Kp & 450.7 & 30.0 & 0.2 & -7.0 & 10.0 & 6.1 & 0.1  \\ 
2008-01-21 & 13:03:12.94 & Keck & NIRC2 & Kp & 417.5 & 33.3 & -1.8 & -6.4 & 10.0 & 6.0 & 0.1  \\ 
2008-01-21 & 13:44:29.32 & Keck & NIRC2 & Kp & 401.7 & 29.8 & 3.1 & -11.5 & 10.0 & 6.4 & 0.3  \\ 
2008-01-21 & 14:26:20.36 & Keck & NIRC2 & Kp & 376.6 & 36.6 & -0.0 & -6.2 & 10.0 & 6.1 & 0.1  \\ 
2008-01-21 & 15:46:10.54 & Keck & NIRC2 & Kp & 331.2 & 27.4 & -0.9 & -17.8 & 10.0 & 6.0 & 0.8  \\ 
2008-01-21 & 16:15:07.30 & Keck & NIRC2 & Kp & 309.8 & 34.8 & -5.4 & -11.2 & 10.0 & 6.6 & 0.2  \\ 
2009-06-07 & 10:04:40.24 & Keck & NIRC2 & H & 519.0 & -199.5 & 8.5 & -14.9 & 10.0 & 6.1 & 0.2  \\ 
2009-06-07 & 10:08:01.37 & Keck & NIRC2 & H & 516.3 & -198.6 & 7.8 & -13.5 & 10.0 & 6.2 & 0.1  \\ 
2009-06-07 & 10:10:56.01 & Keck & NIRC2 & H & 514.3 & -200.6 & 7.5 & -15.1 & 10.0 & 6.2 & 0.1  \\ 
2010-08-15 & 08:37:36.90 & Gemini & NIRI & Kp & 684.0 & -289.9 & 16.8 & -8.1 & 21.9 & 5.8 & 0.0  \\ 
2010-08-23 & 05:24:10.51 & VLT & NACO & Ks & 9.1 & -383.6 & 15.8 & 5.8 & 13.3 & 6.7 & 0.2  \\ 
2010-08-25 & 08:20:44.65 & Gemini & NIRI & Kp & 400.7 & 289.0 & 13.5 & -2.3 & 21.9 & 5.3 & 0.1  \\ 
2010-08-25 & 08:23:50.40 & Gemini & NIRI & Kp & 408.9 & 294.9 & 19.2 & 4.6 & 21.9 & 6.0 & 0.2  \\ 
2010-08-25 & 08:28:19.21 & Gemini & NIRI & Kp & 410.6 & 289.0 & 17.4 & 0.2 & 21.9 & 5.8 & 0.1  \\ 
2010-08-28 & 08:12:04.40 & Gemini & NIRI & Kp & -422.1 & 372.0 & -10.8 & 5.3 & 21.9 & 5.7 & 0.2  \\ 
2010-08-28 & 08:21:40.51 & Gemini & NIRI & Kp & -407.5 & 374.2 & -3.6 & 5.9 & 21.9 & 5.5 & 0.5  \\ 
2010-08-30 & 05:34:11.88 & VLT & NACO & Ks & 354.4 & -372.9 & 26.0 & 5.8 & 13.3 & 6.2 & 0.1  \\ 
2010-08-30 & 05:38:03.19 & VLT & NACO & Ks & 346.7 & -374.7 & 21.5 & 4.4 & 13.3 & 6.1 & 0.1  \\ 
2010-08-30 & 06:19:40.19 & VLT & NACO & Ks & 314.9 & -379.1 & 23.7 & 4.1 & 13.3 & 6.6 & 0.1  \\ 
2010-08-30 & 06:22:24.38 & VLT & NACO & Ks & 305.6 & -376.2 & 16.7 & 7.2 & 13.3 & 6.4 & 0.1  \\ 
2010-09-01 & 08:20:04.50 & Gemini & NIRI & Kp & 41.6 & 377.4 & 0.1 & 4.3 & 21.9 & 5.6 & 0.2  \\ 
2010-09-01 & 08:27:34.01 & Gemini & NIRI & Kp & 49.7 & 378.6 & 1.4 & 6.5 & 21.9 & 5.8 & 0.3  \\ 
2010-09-02 & 06:33:39.81 & Gemini & NIRI & Kp & 753.2 & -121.7 & 17.4 & -8.9 & 21.9 & 5.7 & 0.0  \\ 
2011-10-07 & 02:38:09.32 & VLT & NACO & H & -712.2 & -63.6 & -7.6 & -18.4 & 13.3 & 6.3 & 0.1  \\ 
2011-11-06 & 02:03:27.45 & VLT & NACO & H & 381.0 & -14.6 & 10.1 & 7.5 & 13.3 & 6.4 & 0.1  \\ 
2011-11-08 & 03:05:48.68 & VLT & NACO & H & -604.7 & 8.5 & -6.6 & -0.9 & 13.3 & 5.8 & 0.0  \\ 
2011-11-10 & 01:05:07.29 & VLT & NACO & H & 685.4 & 11.4 & 18.1 & 15.0 & 13.3 & 6.6 & 0.1  \\ 
2011-11-16 & 01:07:15.92 & VLT & NACO & H & -680.8 & -40.4 & 1.4 & -12.8 & 13.3 & 5.8 & 0.2  \\ 
2011-11-20 & 01:06:25.27 & VLT & NACO & H & -355.8 & -30.9 & 9.9 & 8.2 & 13.3 & 6.4 & 0.0  \\ 
2011-12-15 & 04:55:52.51 & Keck & NIRC2 & Kp & -572.8 & -38.9 & 5.0 & -4.4 & 10.0 & 5.4 & 0.0  \\ 
2011-12-15 & 05:06:38.13 & Keck & NIRC2 & H & -570.2 & -39.8 & 3.6 & -4.8 & 10.0 & 5.7 & 0.0  \\ 
2011-12-15 & 05:18:35.71 & Keck & NIRC2 & J & -568.3 & -41.4 & 0.9 & -5.8 & 10.0 & 6.1 & 0.1  \\ 
2011-12-15 & 05:29:27.71 & Keck & NIRC2 & H & -566.0 & -44.0 & -1.0 & -7.8 & 10.0 & 6.0 & 0.1  \\ 
2011-12-15 & 06:28:28.18 & Keck & NIRC2 & H & -537.9 & -47.6 & 2.1 & -8.6 & 10.0 & 5.9 & 0.0  \\ 
2011-12-16 & 04:30:16.20 & Keck & NIRC2 & H & 396.0 & -38.2 & 20.0 & -0.2 & 10.0 & 6.2 & 0.1  \\ 
2011-12-16 & 04:36:40.46 & Keck & NIRC2 & J & 399.5 & -37.1 & 19.4 & 0.6 & 10.0 & 6.3 & 0.1  \\ 
2011-12-16 & 06:01:26.41 & Keck & NIRC2 & H & 446.7 & -27.5 & 15.0 & 6.0 & 10.0 & 6.2 & 0.0  \\ 
2011-12-17 & 06:16:36.73 & Keck & NIRC2 & H & 420.8 & 67.0 & 15.1 & 17.4 & 10.0 & 5.9 & 0.0  \\ 
2011-12-17 & 06:21:36.41 & Keck & NIRC2 & J & 419.1 & 62.7 & 16.4 & 12.9 & 10.0 & 6.0 & 0.0  \\ 
2017-07-10 & 09:45:08.43 & VLT & IRDIS & B\_Y & -292.0 & -118.4 & -17.3 & 7.2 & 12.2 & 6.0 & 0.1  \\ 
2017-07-10 & 09:48:13.66 & VLT & IRDIS & B\_Y & -295.1 & -116.1 & -18.0 & 9.1 & 12.2 & 6.0 & 0.1  \\ 
2017-07-10 & 09:51:17.41 & VLT & IRDIS & B\_Y & -297.8 & -115.7 & -18.2 & 9.2 & 12.2 & 6.0 & 0.0  \\ 
2017-07-10 & 09:59:30.00 & VLT & IRDIS & B\_Y & -306.0 & -112.3 & -19.8 & 11.8 & 12.2 & 5.9 & 0.2  \\ 
2017-07-10 & 10:02:34.87 & VLT & IRDIS & B\_Y & -306.7 & -110.5 & -18.0 & 13.3 & 12.2 & 5.9 & 0.1  \\ 
2017-07-10 & 10:05:39.91 & VLT & IRDIS & B\_Y & -310.5 & -108.0 & -19.4 & 15.4 & 12.2 & 5.8 & 0.1  \\ 
2017-07-10 & 10:08:44.46 & VLT & IRDIS & B\_Y & -313.9 & -111.6 & -20.3 & 11.5 & 12.2 & 5.8 & 0.1  \\ 
2017-07-22 & 09:38:10.25 & VLT & IRDIS & B\_Y & -628.0 & 136.8 & -10.4 & 10.7 & 12.2 & 5.9 & 0.0  \\ 
2017-07-22 & 09:41:15.00 & VLT & IRDIS & B\_Y & -623.6 & 137.0 & -7.8 & 10.6 & 12.2 & 6.0 & 0.1  \\ 
2017-07-22 & 09:47:24.79 & VLT & IRDIS & B\_Y & -619.7 & 137.6 & -7.3 & 10.4 & 12.2 & 6.0 & 0.0  \\ 
2017-07-22 & 09:55:38.44 & VLT & IRDIS & B\_Y & -612.8 & 138.4 & -5.2 & 10.1 & 12.2 & 6.0 & 0.1  \\ 
2018-10-15 & 08:05:02.05 & VLT & ZIMPOL & R & 727.6 & 29.4 & 8.6 & -10.9 & 10.8 & 7.8 & 0.5  \\ 
2018-10-15 & 08:09:38.63 & VLT & ZIMPOL & R & 727.3 & 31.2 & 7.8 & -10.5 & 10.8 & 7.8 & 0.0  \\ 
2018-10-15 & 08:14:15.36 & VLT & ZIMPOL & R & 729.9 & 30.8 & 10.0 & -12.3 & 10.8 & 8.0 & 0.0  \\ 
2018-10-15 & 08:18:50.18 & VLT & ZIMPOL & R & 729.4 & 33.1 & 9.0 & -11.4 & 10.8 & 7.8 & 0.8  \\ 
2018-10-15 & 08:23:25.77 & VLT & ZIMPOL & R & 734.5 & 34.7 & 13.7 & -11.1 & 10.8 & 7.9 & 2.1  \\ 
2018-10-15 & 08:42:22.72 & VLT & ZIMPOL & R & 729.2 & 41.4 & 6.8 & -10.1 & 10.8 & 7.6 & 0.4  \\ 
2018-10-15 & 08:46:57.30 & VLT & ZIMPOL & R & 729.7 & 42.1 & 7.0 & -10.8 & 10.8 & 7.7 & 0.2  \\ 
2018-10-15 & 08:51:31.29 & VLT & ZIMPOL & R & 729.9 & 42.8 & 7.0 & -11.4 & 10.8 & 7.8 & 0.7  \\ 
2018-10-15 & 08:56:04.98 & VLT & ZIMPOL & R & 727.5 & 44.9 & 4.3 & -10.7 & 10.8 & 8.1 & 0.2  \\ 
2018-10-15 & 09:00:40.47 & VLT & ZIMPOL & R & 728.1 & 45.8 & 4.7 & -11.1 & 10.8 & 7.9 & 0.2  \\ 
2018-10-19 & 04:26:15.97 & VLT & ZIMPOL & R & 715.0 & 114.2 & 2.4 & -14.4 & 10.8 & 8.2 & 0.2  \\ 
2018-10-19 & 04:30:52.19 & VLT & ZIMPOL & R & 717.8 & 113.0 & 6.2 & -16.9 & 10.8 & 8.2 & 0.2  \\ 
2018-10-19 & 04:35:29.59 & VLT & ZIMPOL & R & 717.1 & 113.1 & 6.5 & -18.0 & 10.8 & 8.4 & 0.4  \\ 
2018-10-19 & 04:40:04.50 & VLT & ZIMPOL & R & 715.9 & 114.5 & 6.3 & -17.8 & 10.8 & 8.4 & 0.2  \\ 
2018-10-19 & 04:44:38.86 & VLT & ZIMPOL & R & 715.7 & 116.5 & 7.1 & -17.0 & 10.8 & 8.3 & 0.3  \\ 
2018-11-12 & 05:57:28.60 & VLT & ZIMPOL & R & -425.4 & -257.6 & -7.9 & 0.4 & 10.8 & 8.4 & 0.1  \\ 
2018-11-12 & 06:02:03.30 & VLT & ZIMPOL & R & -421.8 & -257.6 & -7.8 & 0.9 & 10.8 & 7.9 & 0.1  \\ 
2018-11-12 & 06:06:38.46 & VLT & ZIMPOL & R & -418.3 & -258.7 & -7.8 & 0.3 & 10.8 & 7.9 & 0.1  \\ 
2018-11-12 & 06:11:12.56 & VLT & ZIMPOL & R & -413.6 & -258.7 & -6.6 & 0.7 & 10.8 & 8.0 & 0.1  \\ 
2018-11-12 & 06:15:47.66 & VLT & ZIMPOL & R & -409.8 & -259.8 & -6.2 & 0.1 & 10.8 & 7.6 & 0.0  \\ 
2018-11-12 & 07:02:46.55 & VLT & ZIMPOL & R & -367.8 & -265.8 & -1.5 & -1.9 & 10.8 & 8.3 & 0.1  \\ 
2018-11-12 & 07:07:20.61 & VLT & ZIMPOL & R & -363.7 & -265.4 & -1.0 & -1.2 & 10.8 & 8.1 & 0.2  \\ 
2018-11-12 & 07:11:55.72 & VLT & ZIMPOL & R & -360.4 & -264.8 & -1.4 & -0.2 & 10.8 & 8.3 & 0.4  \\ 
2018-11-12 & 07:16:30.83 & VLT & ZIMPOL & R & -355.7 & -270.7 & -0.3 & -5.8 & 10.8 & 9.3 & 0.6  \\ 
2018-11-25 & 04:31:28.34 & VLT & ZIMPOL & R & 151.3 & 274.7 & 4.5 & 3.7 & 10.8 & 7.1 & 0.1  \\ 
2018-11-25 & 04:36:04.60 & VLT & ZIMPOL & R & 145.8 & 273.8 & 3.0 & 2.8 & 10.8 & 7.1 & 0.1  \\ 
2018-11-25 & 04:40:40.42 & VLT & ZIMPOL & R & 140.7 & 274.9 & 2.0 & 4.0 & 10.8 & 7.1 & 0.1  \\ 
2018-11-25 & 04:45:15.95 & VLT & ZIMPOL & R & 136.4 & 275.3 & 1.7 & 4.4 & 10.8 & 6.9 & 0.1  \\ 
2018-11-25 & 04:49:51.17 & VLT & ZIMPOL & R & 132.7 & 274.8 & 2.1 & 4.0 & 10.8 & 6.9 & 0.0  \\ 
2018-11-26 & 02:31:28.29 & VLT & ZIMPOL & R & -757.2 & -7.5 & -9.9 & 10.9 & 10.8 & 7.7 & 0.2  \\ 
2018-11-26 & 02:36:04.72 & VLT & ZIMPOL & R & -757.6 & -8.6 & -9.6 & 11.3 & 10.8 & 7.7 & 0.1  \\ 
2018-11-26 & 02:40:41.57 & VLT & ZIMPOL & R & -758.5 & -9.8 & -9.8 & 11.6 & 10.8 & 7.9 & 0.1  \\ 
2018-11-26 & 02:45:16.87 & VLT & ZIMPOL & R & -760.0 & -12.3 & -10.6 & 10.6 & 10.8 & 7.9 & 0.3  \\ 
2018-11-26 & 02:49:51.73 & VLT & ZIMPOL & R & -759.9 & -13.9 & -9.9 & 10.4 & 10.8 & 8.1 & 0.4  \\ 
2018-11-29 & 04:23:20.66 & VLT & ZIMPOL & R & -315.4 & 217.1 & -1.8 & 5.4 & 10.8 & 8.7 & 0.1  \\ 
2018-11-29 & 04:27:55.88 & VLT & ZIMPOL & R & -319.0 & 216.5 & -1.7 & 5.7 & 10.8 & 8.5 & 0.3  \\ 
2018-11-29 & 04:32:31.55 & VLT & ZIMPOL & R & -321.7 & 216.1 & -0.7 & 6.2 & 10.8 & 7.8 & 0.3  \\ 
2018-11-29 & 04:37:06.06 & VLT & ZIMPOL & R & -325.2 & 214.5 & -0.5 & 5.5 & 10.8 & 8.0 & 0.4  \\ 
2018-11-29 & 04:41:40.71 & VLT & ZIMPOL & R & -329.7 & 213.9 & -1.3 & 5.8 & 10.8 & 8.0 & 0.3  \\ 
  \end{longtable}
\end{center}

\begin{center}
  \begin{longtable}{cclllrrrrrrr}
  \caption[Astrometry of Remus]{Astrometry of Sylvia's satellite Remus.
    Date, mid-observing time (UTC), telescope, camera, filter, 
    astrometry ($X$ is aligned with Right Ascension, and $Y$ with Declination, and 
    $o$ and $c$ indices stand for observed and computed positions),
    uncertainty ($\sigma$), 
    and photometry (magnitude difference $\Delta M$ with uncertainty $\delta M$).
    \label{tab:sat2}
  }\\

    \hline\hline
     Date & UTC & Tel. & Cam. & Filter &
     \multicolumn{1}{c}{$X_o$} &
     \multicolumn{1}{c}{$Y_o$} &
     \multicolumn{1}{c}{$X_{o-c}$} &
     \multicolumn{1}{c}{$Y_{o-c}$} &
     \multicolumn{1}{c}{$\sigma$} &
     \multicolumn{1}{c}{$\Delta M$} &
     \multicolumn{1}{c}{$\delta M$} \\
    &&&&& 
     \multicolumn{1}{c}{(mas)} & \multicolumn{1}{c}{(mas)} & 
     \multicolumn{1}{c}{(mas)} & \multicolumn{1}{c}{(mas)} & 
     \multicolumn{1}{c}{(mas)} & 
     \multicolumn{1}{c}{(mag)} & \multicolumn{1}{c}{(mag)}  \\ 
    \hline
    \endfirsthead

    \multicolumn{11}{c}{{\tablename\ \thetable{} -- continued from previous page}} \\ 
    \hline\hline
     Date & UTC & Tel. & Cam. & Filter &
     \multicolumn{1}{c}{$X_o$} &
     \multicolumn{1}{c}{$Y_o$} &
     \multicolumn{1}{c}{$X_{o-c}$} &
     \multicolumn{1}{c}{$Y_{o-c}$} &
     \multicolumn{1}{c}{$\sigma$} &
     \multicolumn{1}{c}{$\Delta M$} &
     \multicolumn{1}{c}{$\delta M$} \\
    &&&&& 
     \multicolumn{1}{c}{(mas)} & \multicolumn{1}{c}{(mas)} & 
     \multicolumn{1}{c}{(mas)} & \multicolumn{1}{c}{(mas)} & 
     \multicolumn{1}{c}{(mas)} & 
     \multicolumn{1}{c}{(mag)} & \multicolumn{1}{c}{(mag)}  \\ 
    \hline
    \endhead

    \hline \multicolumn{11}{r}{{Continued on next page}} \\ \hline
    \endfoot

    \hline
    \endlastfoot
2003-06-05 & 12:54:41.82 & Keck & NIRC2 & Ks & 396.8 & 5.9 & 22.0 & -0.6 & 10.0 & 7.6 & 0.2  \\ 
2003-08-14 & 07:02:32.56 & Keck & NIRC2 & H & -262.5 & -64.0 & 8.9 & 11.0 & 10.0 & 7.0 & 0.2  \\ 
2003-08-14 & 07:10:03.39 & Keck & NIRC2 & H & -285.9 & -76.2 & -9.6 & -5.2 & 10.0 & 8.5 & 1.2  \\ 
2004-08-10 & 07:17:15.21 & VLT & NACO & Ks & 407.6 & -1.8 & -3.3 & 2.2 & 13.3 & 8.0 & 1.0  \\ 
2004-09-01 & 05:54:49.70 & VLT & NACO & Ks & 238.9 & -104.4 & 3.5 & -0.3 & 13.3 & 9.7 & 1.2  \\ 
2004-09-01 & 06:03:51.58 & VLT & NACO & H & 225.6 & -104.4 & 0.7 & 0.9 & 13.3 & 10.2 & 1.2  \\ 
2004-09-03 & 07:08:59.76 & VLT & NACO & Ks & -213.8 & 100.0 & -4.5 & -7.0 & 13.3 & 6.9 & 0.1  \\ 
2004-09-04 & 08:41:05.05 & VLT & NACO & Ks & -424.6 & -14.4 & -13.7 & -4.1 & 13.3 & 10.7 & 1.2  \\ 
2004-09-05 & 04:09:23.01 & VLT & NACO & Ks & 411.4 & -54.0 & 8.4 & 5.5 & 13.3 & 10.7 & 1.2  \\ 
2004-09-07 & 02:36:18.09 & VLT & NACO & Ks & -437.9 & 0.0 & -13.3 & -8.5 & 13.3 & 9.9 & 1.2  \\ 
2004-11-02 & 07:34:31.11 & Gemini & NIRI & Kp & 312.8 & 5.1 & -2.1 & -14.0 & 21.9 & 5.2 & 0.0  \\ 
2005-12-21 & 08:47:19.69 & Gemini & NIRI & Ks & 331.5 & -17.8 & -9.3 & 1.3 & 21.9 & 7.6 & 1.2  \\ 
2006-01-01 & 10:43:05.31 & Gemini & NIRI & Ks & 286.8 & 90.5 & 15.0 & -13.7 & 21.9 & 8.3 & 1.2  \\ 
2008-01-21 & 11:24:55.26 & Keck & NIRC2 & Kp & -265.9 & -7.1 & -1.5 & 7.8 & 10.0 & 6.8 & 0.1  \\ 
2008-01-21 & 11:56:28.79 & Keck & NIRC2 & Kp & -251.1 & -5.1 & -2.9 & 11.5 & 10.0 & 7.1 & 1.7  \\ 
2010-08-28 & 08:12:04.40 & Gemini & NIRI & Kp & -352.2 & -68.1 & -4.0 & -19.2 & 21.9 & -0.2 & 0.2  \\ 
2010-08-28 & 08:21:40.51 & Gemini & NIRI & Kp & -350.5 & -65.8 & 3.0 & -23.0 & 21.9 & 6.8 & 1.2  \\ 
2010-09-01 & 08:20:04.50 & Gemini & NIRI & Kp & -284.2 & -110.4 & -6.1 & -8.2 & 21.9 & 5.5 & 0.9  \\ 
2010-09-01 & 08:27:34.01 & Gemini & NIRI & Kp & -284.9 & -109.5 & -0.5 & -11.5 & 21.9 & 7.2 & 1.2  \\ 
2010-09-02 & 06:33:39.81 & Gemini & NIRI & Kp & 374.2 & -128.5 & 22.3 & -12.5 & 21.9 & 7.2 & 1.3  \\ 
2011-10-07 & 02:38:09.32 & VLT & NACO & H & -396.6 & -21.3 & 7.1 & -0.2 & 13.3 & 7.9 & 0.1  \\ 
2011-11-08 & 03:05:48.68 & VLT & NACO & H & 212.0 & 32.4 & -1.0 & 8.7 & 13.3 & 8.2 & 0.0  \\ 
2011-11-10 & 01:05:07.29 & VLT & NACO & H & -358.1 & -26.7 & -2.5 & -9.8 & 13.3 & 0.6 & 1.2  \\ 
2011-11-16 & 01:07:15.92 & VLT & NACO & H & 392.8 & 18.7 & 4.1 & 11.7 & 13.3 & 6.8 & 2.2  \\ 
2011-12-15 & 04:55:52.51 & Keck & NIRC2 & Kp & -337.4 & -2.8 & 1.2 & 0.7 & 10.0 & 6.9 & 0.6  \\ 
2011-12-15 & 05:06:38.13 & Keck & NIRC2 & H & -337.2 & 1.4 & 1.7 & 6.0 & 10.0 & 7.3 & 0.3  \\ 
2011-12-15 & 05:18:35.71 & Keck & NIRC2 & J & -337.4 & -4.4 & 1.5 & 1.4 & 10.0 & 8.3 & 0.1  \\ 
2011-12-15 & 05:29:27.71 & Keck & NIRC2 & H & -338.1 & -5.3 & 0.2 & 1.6 & 10.0 & 7.8 & 0.4  \\ 
2011-12-15 & 06:28:28.18 & Keck & NIRC2 & H & -330.6 & -21.4 & -2.3 & -8.7 & 10.0 & 8.4 & 1.2  \\ 
2011-12-17 & 06:16:36.73 & Keck & NIRC2 & H & 356.2 & 8.6 & 20.4 & 1.3 & 10.0 & 7.8 & 1.2  \\ 
2011-12-17 & 06:21:36.41 & Keck & NIRC2 & J & 345.9 & 8.3 & 10.5 & 0.5 & 10.0 & 7.6 & 1.2  \\ 
2017-07-10 & 09:42:03.69 & VLT & IRDIS & B\_Y & 667.5 & 91.4 & 12.4 & -20.7 & 12.2 & 0.9 & 0.0  \\ 
2017-07-10 & 09:45:08.43 & VLT & IRDIS & B\_Y & 380.5 & -28.4 & -2.6 & -13.9 & 12.2 & 7.1 & 0.1  \\ 
2017-07-10 & 09:48:13.66 & VLT & IRDIS & B\_Y & 380.5 & -28.3 & -2.8 & -13.1 & 12.2 & 7.1 & 0.1  \\ 
2017-07-10 & 09:51:17.41 & VLT & IRDIS & B\_Y & 380.5 & -27.4 & -2.9 & -11.5 & 12.2 & 7.1 & 0.1  \\ 
2017-07-10 & 09:59:30.00 & VLT & IRDIS & B\_Y & 379.7 & -29.4 & -3.9 & -11.6 & 12.2 & 7.2 & 0.1  \\ 
2017-07-10 & 10:02:34.87 & VLT & IRDIS & B\_Y & 379.6 & -29.3 & -3.9 & -10.8 & 12.2 & 7.1 & 0.1  \\ 
2017-07-10 & 10:05:39.91 & VLT & IRDIS & B\_Y & 382.2 & -28.8 & -1.3 & -9.6 & 12.2 & 7.2 & 0.1  \\ 
2017-07-10 & 10:08:44.46 & VLT & IRDIS & B\_Y & 378.8 & -28.8 & -4.6 & -8.9 & 12.2 & 7.0 & 0.1  \\ 
2017-07-22 & 09:38:10.25 & VLT & IRDIS & B\_Y & 203.7 & 61.2 & 2.2 & 0.2 & 12.2 & 7.5 & 0.3  \\ 
2017-07-22 & 09:41:15.00 & VLT & IRDIS & B\_Y & 206.3 & 56.9 & 1.4 & -3.5 & 12.2 & 7.3 & 0.9  \\ 
2017-07-22 & 09:47:24.79 & VLT & IRDIS & B\_Y & 209.3 & 56.4 & -2.4 & -2.9 & 12.2 & 7.7 & 0.0  \\ 
2017-07-22 & 09:55:38.44 & VLT & IRDIS & B\_Y & 229.5 & 57.4 & 8.8 & -0.3 & 12.2 & 7.5 & 0.9  \\ 
2018-11-12 & 05:57:28.60 & VLT & ZIMPOL & R & 290.5 & -70.6 & 1.4 & -3.8 & 10.8 & 10.1 & 0.1  \\ 
2018-11-12 & 06:02:03.30 & VLT & ZIMPOL & R & 293.7 & -68.9 & 0.6 & -3.9 & 10.8 & 9.0 & 0.5  \\ 
2018-11-12 & 06:06:38.46 & VLT & ZIMPOL & R & 298.7 & -66.4 & 1.7 & -3.3 & 10.8 & 9.0 & 0.6  \\ 
2018-11-12 & 06:11:12.56 & VLT & ZIMPOL & R & 304.2 & -65.0 & 3.4 & -3.8 & 10.8 & 9.0 & 0.3  \\ 
2018-11-12 & 06:15:47.66 & VLT & ZIMPOL & R & 307.5 & -63.2 & 3.0 & -4.0 & 10.8 & 8.8 & 0.1  \\ 
2018-11-12 & 07:02:46.55 & VLT & ZIMPOL & R & 344.5 & -40.1 & 5.1 & -1.4 & 10.8 & 9.0 & 0.1  \\ 
2018-11-12 & 07:07:20.61 & VLT & ZIMPOL & R & 343.9 & -38.1 & 1.5 & -1.4 & 10.8 & 8.9 & 0.3  \\ 
2018-11-12 & 07:11:55.72 & VLT & ZIMPOL & R & 349.5 & -36.6 & 4.3 & -1.9 & 10.8 & 9.2 & 1.9  \\ 
2018-11-12 & 07:16:30.83 & VLT & ZIMPOL & R & 349.0 & -50.3 & 0.9 & -17.7 & 10.8 & 10.3 & 1.2  \\ 
2018-11-25 & 04:31:28.34 & VLT & ZIMPOL & R & -354.4 & 32.3 & -7.6 & -0.6 & 10.8 & 9.3 & 0.2  \\ 
2018-11-25 & 04:36:04.60 & VLT & ZIMPOL & R & -357.8 & 30.3 & -8.3 & -0.5 & 10.8 & 8.9 & 0.1  \\ 
2018-11-25 & 04:40:40.42 & VLT & ZIMPOL & R & -361.5 & 29.3 & -9.5 & 0.5 & 10.8 & 8.9 & 0.1  \\ 
2018-11-25 & 04:45:15.95 & VLT & ZIMPOL & R & -364.5 & 27.7 & -10.0 & 1.0 & 10.8 & 8.5 & 0.1  \\ 
2018-11-25 & 04:49:51.17 & VLT & ZIMPOL & R & -366.7 & 25.2 & -9.8 & 0.6 & 10.8 & 8.4 & 0.1  \\ 
2018-11-26 & 02:31:28.29 & VLT & ZIMPOL & R & 329.8 & 107.8 & 10.3 & -5.2 & 10.8 & 9.6 & 0.4  \\ 
2018-11-26 & 02:36:04.72 & VLT & ZIMPOL & R & 327.6 & 108.9 & 11.5 & -5.4 & 10.8 & 10.0 & 0.2  \\ 
2018-11-26 & 02:40:41.57 & VLT & ZIMPOL & R & 324.1 & 112.6 & 11.5 & -3.1 & 10.8 & 10.6 & 0.6  \\ 
2018-11-26 & 02:45:16.87 & VLT & ZIMPOL & R & 318.7 & 114.0 & 9.6 & -2.9 & 10.8 & 10.3 & 1.9  \\ 
2018-11-26 & 02:49:51.73 & VLT & ZIMPOL & R & 307.5 & 113.1 & 2.0 & -5.1 & 10.8 & 11.0 & 0.4  \\ 
2018-11-29 & 04:23:20.66 & VLT & ZIMPOL & R & -261.8 & 72.6 & 3.0 & -5.7 & 10.8 & 7.6 & 1.2  \\ 
2018-11-29 & 04:27:55.88 & VLT & ZIMPOL & R & -264.8 & 71.9 & 4.2 & -4.6 & 10.8 & 7.4 & 1.2  \\ 
2018-11-29 & 04:32:31.55 & VLT & ZIMPOL & R & -269.5 & 69.6 & 3.6 & -5.2 & 10.8 & 9.3 & 0.3  \\ 
2018-11-29 & 04:37:06.06 & VLT & ZIMPOL & R & -269.9 & 64.7 & 7.3 & -8.2 & 10.8 & 7.2 & 0.4  \\ 
2018-11-29 & 04:41:40.71 & VLT & ZIMPOL & R & -278.6 & 66.5 & 2.5 & -4.6 & 10.8 & 10.9 & 0.3  \\ 
  \end{longtable}
\end{center}

\end{appendix}

\end{document}